\documentclass[twocolumn]{aastex63}		         
 
\usepackage{float}
\usepackage{comment}
\usepackage[]{natbib}
\usepackage{placeins}
\usepackage{graphicx}	
\usepackage{multirow}
\usepackage{amsmath}

\shorttitle{BH Activity and Outflows in a Dwarf Galaxy}

\newcounter{species} 
\def\ion#1#2{\hbox{\setcounter{species}{#2}#1\,{\scriptsize\Roman{species}}\relax}}

\def\lsim{\lower0.3em\hbox{$\,\buildrel <\over\sim\,$}}
\def\gsim{\lower0.3em\hbox{$\,\buildrel >\over\sim\,$}}

\usepackage[normalem]{ulem}
\def\crossout#1{}
\def\crossout#1{\sout{#1}}

\received{December 10, 2020}
\revised{January 20, 2020}
\accepted{January 27, 2020}
\begin{document}
\title{Outflows, Shocks and Coronal Line Emission in a Radio-Selected AGN in a Dwarf Galaxy}

\correspondingauthor{Mallory Molina}
\email{mallory.molina@montana.edu}

\author[0000-0001-8440-3613]{Mallory Molina} 
\affil{eXtreme Gravity Institute, Department of Physics, Montana State University, Bozeman, MT 59717, USA}

\author[0000-0001-7158-614X]{Amy E. Reines}
\affil{eXtreme Gravity Institute, Department of Physics, Montana State University, Bozeman, MT 59717, USA}

\author{Jenny E. Greene}
\affiliation{Department of Astrophysical Sciences, Princeton University, Princeton, NJ 08544, USA}

\author[0000-0003-2511-2060]{Jeremy Darling}
\affiliation{Center for Astrophysics and Space Astronomy, Department of Astrophysical and Planetary Sciences, University of Colorado, 389 UCB, Boulder, CO 80309-0389, USA}

\author[0000-0003-4724-1939]{James J. Condon}
\affiliation{National Radio Astronomy Observatory, Charlottesville, VA, 22903, USA}

\begin{abstract}
Massive black holes (BHs) in dwarf galaxies can provide strong constraints on BH seeds, however reliably detecting them is notoriously difficult. High resolution radio observations were recently used to identify accreting massive BHs in nearby dwarf galaxies, with a significant fraction found to be non-nuclear. Here we present the first results of our optical follow-up of these radio-selected active galactic nuclei (AGNs) in dwarf galaxies using integral field unit (IFU) data from Gemini-North.  We focus on the dwarf galaxy J1220+3020, which shows no clear optical AGN signatures in its nuclear SDSS spectrum covering the radio source. With our new IFU data, we confirm the presence of an active BH via the AGN coronal line [\ion{Fe}{10}] and enhanced [\ion{O}{1}] emission coincident with the radio source.  Furthermore, we detect broad H$\alpha$ emission and estimate a BH mass of 
$M_{\rm BH}=10^{4.9}M_\odot$. We compare the narrow emission line ratios to standard BPT diagnostics and shock models. Spatially-resolved BPT diagrams show some AGN signatures, particularly in [\ion{O}{1}]/H$\alpha$, but overall do not unambiguously identify the AGN. A comparison of our data to shock models clearly indicates shocked emission surrounding the AGN. The physical model most consistent with the data is an active BH with a radiatively inefficient accretion flow (RIAF) that both photoionizes and shock-excites the surrounding gas. We conclude that feedback is important in radio-selected BHs in dwarf galaxies, and that radio surveys may probe a population of low accretion-rate BHs in dwarf galaxies that cannot be detected through optical surveys alone.
\end{abstract}
\keywords{active galaxies -- dwarf galaxies -- low-luminosity active galactic nuclei -- radio jets -- black holes}

\section{Introduction}
\label{sec:intro}
Supermassive black holes (BHs) are known to be ubiquitous in the nuclei of massive galaxies \citep[e.g.,][]{Kormendy1995,Kormendy2013}, but their initial formation conditions have been lost to merger-driven growth over cosmic time \citep{Volonteri2010,Natarajan2014}. The proposed theories for the formation of the initial BH ``seeds in the early Universe include remnants from Population III stars, which would create BHs with $M_{\rm BH}\sim100M_\odot$ \citep{Bromm2011}. Alternatively, the direct collapse of gas  \citep{Loeb1994,Begelman2006,Lodato2006,Choi2015}, or collisions in dense star clusters \citep{Portegies2004,Devecchi2009,Davies2011,Lupi2014,Stone2017} would form much larger initial BH seeds, with masses $M_{\rm BH}\sim10^3$--$10^5M_\odot$. 

While the first BH ``seeds'' at high redshift are too small and faint to currently detect, nearby dwarf galaxies can place strong constraints on BH seed masses \citep[see][and references therein]{reines2016, Greene2020}. Dwarf galaxies are known to have relatively quiet merger histories compared to more massive galaxies \citep{Bellovary2011}, and recent work has shown that dwarf galaxies may experience stunted BH growth due to supernova feedback \citep{Angles2017,Habouzit2017}. Thus, if there is a BH in a dwarf galaxy, its mass ($M_{\rm BH}\lesssim10^5$) is expected to be relatively close to the initial seed mass. 

Unfortunately detecting BHs in dwarf galaxies is difficult, due to their smaller masses and lower luminosities. In principle, there are numerous ways to identify accreting BHs, in many different wavelength regimes \citep[see][for a complete review]{Ho2008}. In the optical regime, broad H$\alpha$ emission and/or narrow-line ratios consistent with AGN photoionization have been used to identify low-mass AGNs in dwarf galaxies \citep[e.g.,][]{reines2013}. This work often relies on the Baldwin, Phillips and Terlevich diagrams \citep[BPT diagrams;][]{Baldwin1981}, which differentiate between the harder spectral energy distributions (SEDs) created by AGNs and those of star forming regions. However, these diagrams are not as effective at identifying low-luminosity AGNs (LLAGNs), and low ionization nuclear emission regions \citep[LINERs; see][for a review]{Ho2008,Kewley2019}. LINERs with LLAGNs in massive galaxies will have a relatively larger contribution from shocks than higher luminosity AGNs, thus creating a combination of power sources that contribute to the observed emission \citep{Molina2018}. Similarly, radiatively inefficient accretion flows (RIAFs), which require a low accretion rate, have a different SED that could remove any observable broad-line emission \citep{Trump2011}. Furthermore, optical AGN indicators may be diluted by the host galaxy in these low-metallicity systems, even without a significant amount of ongoing star formation \citep{Moran2002,Groves2006,Stasinska2006,Cann2019}. Given the low luminosity nature of the BHs in dwarf galaxies, any combination of these scenarios can easily hide AGN activity.

Infrared (IR) data, particularly the {\it Wide-field Infrared Survey Explorer} \citep[WISE;][]{Wright2010} colors are also used to identify AGNs in more massive galaxies \citep{Jarrett2011,Mateos2012,Stern2012}, but can be confused with star formation in dwarf galaxies \citep[][Latimer et al. in prep]{Hainline2016,Condon2019}. Similarly, almost all AGNs, even LLAGNs, produce radio emission \citep[see][and references therein]{Ho2008}, which is not affected by dust attenuation. However, the radio emission from BHs in dwarf galaxies could be similar with that from \ion{H}{2} regions or supernova remnants or supernovae \citep[e.g.,][]{,Reines2008,Chomiuk2009,Johnson2009,aversa2011,Kepley2014,Varenius2019}. Therefore, caution must be used when identifying AGNs in dwarf galaxies using either radio or IR data.

\cite{reines2020} conducted a radio survey using the NSF Karl G. Jansky Very Large Array (VLA), and identified 13 dwarf galaxies with radio emission indicating AGNs, some of which were outside of the nucleus of the galaxy. For all 13 AGN candidates, they proved the observed radio emission was too luminous and compact to be explained by stellar processes, such as thermal \ion{H}{2} regions, and individual or populations of supernovae and supernova remnants \citep[see Section 5 of][]{reines2020}. While \cite{reines2020} found that the AGN candidates showed enhanced [\ion{O}{1}]/H$\alpha$ emission in their Sloan Digital Sky Survey spectra \citep[SDSS][]{Blanton2017}, most of them did not have clear optical signatures of AGN photoionization. Given the SDSS spectra have a 3\arcsec diameter, they have a significant contribution from the host galaxy which could contaminate the AGN signal. Therefore, higher-resolution optical data is crucial to confirm these BH candidates.

\begin{figure}
    \centering
    \includegraphics[width=0.46\textwidth]{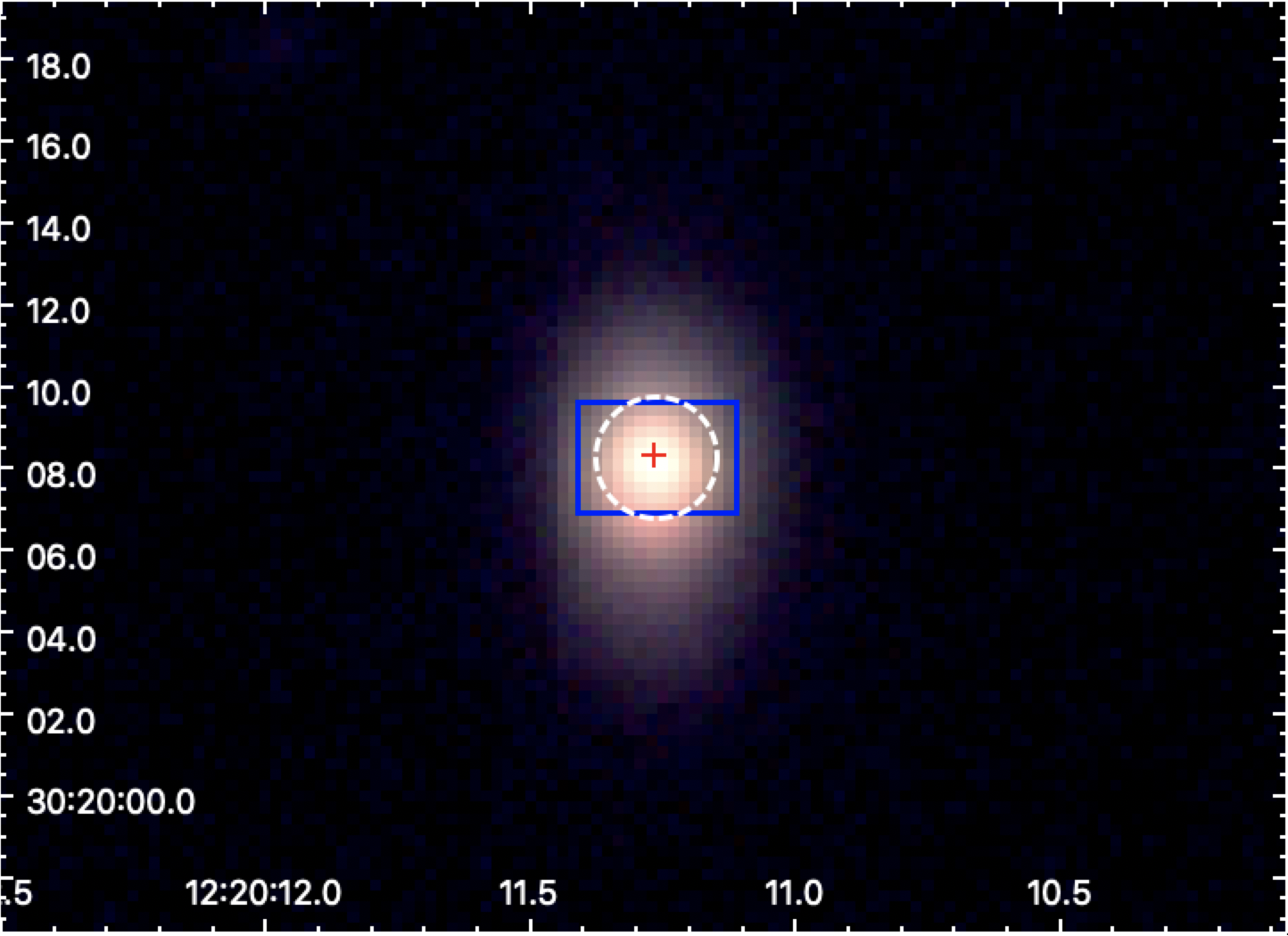}
        \hspace{2pt}\includegraphics[width=0.46\textwidth]{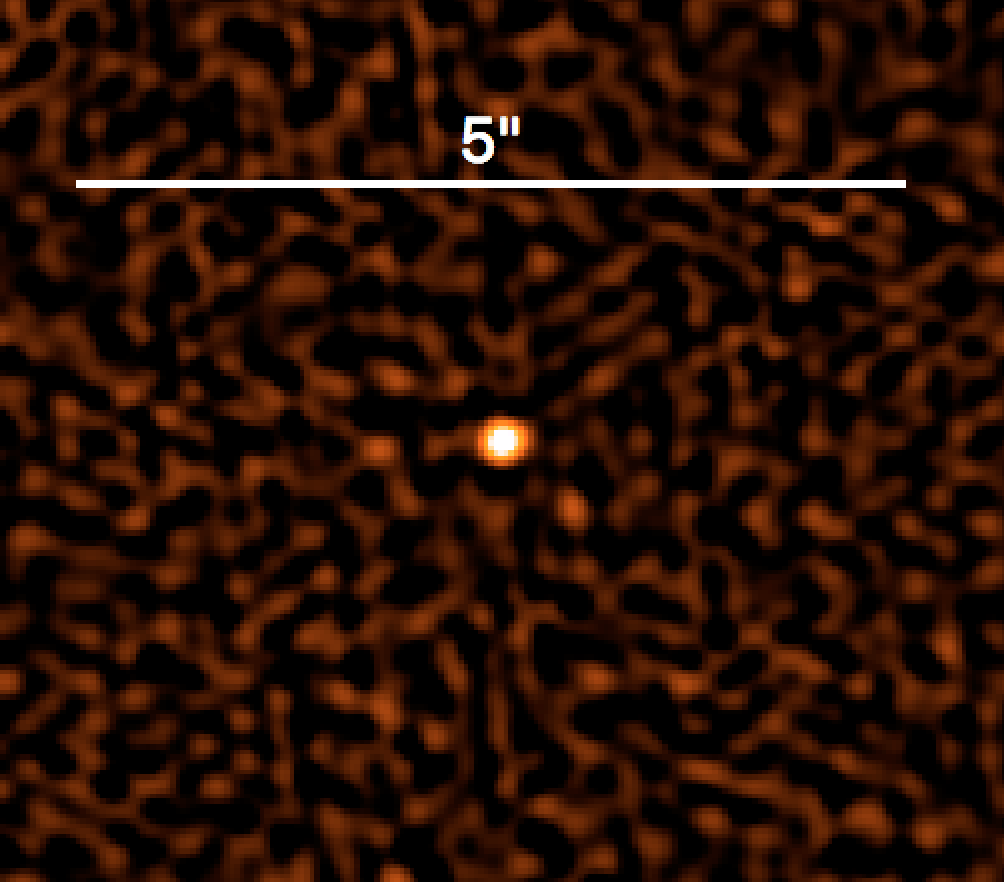}
    \caption{{\it Top: }DECaLS $grz$-band images of dwarf galaxy J1220+3020. The red cross indicates the position of the radio source, the white dashed circle indicates the position of the SDSS fiber, and the blue box is the $2\farcs7\times3\farcs9$ GMOS IFU footprint. {\it Bottom: }9 GHz VLA image of J1220+3020. We show a 5\arcsec white scale bar above the radio source for reference. The emission is consistent with a point source of size $0\farcs18\times0\farcs17$. The properties of J1220+3020 are listed in Table~\ref{table:gprop}.}
    \label{fig:gal}
\end{figure}

In this work, we present the first of the optical integral field unit (IFU) follow-up of the \citet{reines2020} sample taken with the Gemini Multi-Object Spectrograph on Gemini-North (GMOS-N). Our main goals are to confirm the presence of an AGN and study its impact on the host galaxy. In this paper, we present the results for SDSS J122011.26+302008.1 (ID 82 in \citealt{reines2020}), hereafter referred to as J1220+3020. The Dark Energy Camera Legacy Survey \citep[DECaLS;][]{decals} optical image and the VLA radio emission for this galaxy are shown in Figure~\ref{fig:gal}, and the galaxy properties are listed in Table~\ref{table:gprop}. The \cite{reines2020} VLA observation for J1220+3020 was consistent with a point source at a resolution of $\sim0\farcs18$, which corresponds to $\approx100$~pc. The radio source in J1220+3020 is in the nucleus of the galaxy, resides in one of the brighter objects in the sample, and was observed for the full time requested\footnote{Due to the Covid-19 Pandemic, our observing program was not completed.}. We therefore will use J1220+3020 as a case study to test the methodology of our analysis, and compare the properties of radio-selected and optically-selected AGN.

We will make full use of our IFU data by first studying 2--D emission-line maps, especially [\ion{O}{1}]. We will identify any strong morphological features of the gas and discuss their physical implications. After that, we will create resolved 1--D spectra for the radio source and surrounding regions, and compare them to optical diagnostics to determine their dominant excitation mechanism. Given the previously described issues with standard BPT diagrams, we will also explore shock models from \cite{allen2008}, which have been successfully used to explain LINER-like emission in more massive galaxies \citep{Molina2018}.
\begin{deluxetable}{lcl}
  \tablecaption{Properties of J1220+3020\tablenotemark{a}\label{table:gprop}}
\setlength{\tabcolsep}{15pt}
\tablewidth{0pt}
\tablehead{
{Property} &{Value} &{Units}\vspace{-1mm}}
\startdata
{Galaxy R.A.\tablenotemark{b}} & {185.04692} & {degrees}\\
{Galaxy Dec.\tablenotemark{b}} & {30.33562} & {degrees}\\
{Redshift} & {0.0269} & {...}\\
{log(M$_*$/M$_\odot$)} & {9.4} & {...}\\
{$M_g$\tablenotemark{c}} & {$-18.21\pm0.02$} & {mag}\\
{$g-r$\tablenotemark{c}} & {$0.35\pm0.01$} & {mag}\\
{$r_{50}$\tablenotemark{d}} & {1.09} & {kpc}\\
{S\'ersic n} & {4.2} & {...}\\
{$S_{1.4~{\rm GHz}}$\tablenotemark{e}} & {$1.15\pm0.15$} & {mJy}\\
{Radio Source R.A.\tablenotemark{f}} & {185.04694} & {degrees}\\
{Radio Source Dec.\tablenotemark{f}} & {30.33564} & {degrees}\\
{Radio Source Offset\tablenotemark{f}} & {0.1} & {arcsec}\\
{$S_{9~{\rm GHz}}$} & {$397\pm24$} & {$\mu$Jy}\\
{log(L$_{9~{\rm GHz}}$)} & {$20.83\pm0.03$} & {W~Hz$^{-1}$}\\
{$\alpha^{10.65}_{9}$} & {$-0.7\pm0.6$\tablenotemark{g}} &{...}\\
{$\alpha^{1.4}_{9}$} & {$-0.57\pm0.08$\tablenotemark{g}} &{...}\\
{Point Source?\tablenotemark{h}} & {True} &{...}\\
\enddata
\tablenotetext{a}{All reported values are from \cite{reines2020}, and assume $h=0.73$.}
\tablenotetext{b}{The coordinates for the galactic nucleus as defined by the NASA-Sloan Atlas \citep[NSA;][]{Blanton2011}.}
\tablenotetext{c}{Magnitudes are both K-corrected and corrected for foreground Galactic extinction.}
\tablenotetext{d}{$r_{50}$ is the Petrosian 50\% radius.}
\tablenotetext{e}{The 1.4~GHz detection is from the VLA Faint Images of the Radio Sky at Twenty centimeters (FIRST) Survey, with errors calculated using the procedure described in \cite{Condon1997}.}
\tablenotetext{f}{The position of the 9~GHz radio emission as detected by \cite{reines2020}. The ``Radio Source Offset'' is the difference between the NSA galactic nucleus coordinates and the radio source coordinates.}
\tablenotetext{g}{The spectral indices assume $S_\nu\propto\nu^\alpha$. We also present the new calculation of $\alpha^{1.4}_{9}$, but note that the two observations were taken years apart, and that the 1.4~GHz flux measurement could have contamination from star formation in the host galaxy.}
\tablenotetext{h}{The radio emission in  J1220+3020 is consistent with a point source, where the fitted 2--D Gaussian model has a major axis full-width at half-maximum (FWHM) of 0\farcs18 and a minor axis FWHM of 0\farcs17.}
\end{deluxetable}

In Section~\ref{sec:data} we describe the observations and data reduction process. The 2--D spatial emission is presented in Section~\ref{sec:2dem}, and the resolved 1--D emission, including emission-line fitting, is discussed in Section~\ref{sec:1dres}. We compare our data to BPT diagrams and shock models in Section~\ref{sec:mod_comp}. We consider these results and discuss potential physical models for the system in Section~\ref{sec:discussion} and summarize our findings and conclusions in Section~\ref{sec:summary}. In this paper, we assume a $\Lambda\textrm{CDM}$ cosmology with $\Omega_{\rm m}=0.3$, $\Omega_{\Lambda}=0.7$ and H$_0=70$~km~s$^{-1}$~Mpc$^{-1}$. 

\section{Observations and Data Reduction}
\label{sec:data}
\subsection{Description of Observations}
\label{ssec:obs}
The GMOS-N/IFU data for J1220+3020 were taken between February and July 2020. The GMOS-N/IFU uses a hexagonal lenslet array with projected diameters of $0\farcs2$ to fully sample the $7\arcsec\times5\arcsec$ field of view (FOV). Each lenslet is coupled to a fiber, which redirects the light to GMOS. The IFU can be used in either one-slit or two-slit mode. The one-slit mode will record roughly half the number of fibers as the two-slit mode, creating a $2\farcs7\times3\farcs9$ FOV, but will provide extended wavelength coverage. Given our interest creating BPT diagrams (i.e., from H$\beta$ to [\ion{S}{2}]$\lambda\lambda6716,6732$), we employed the one-slit (IFU-R) mode and the B600 filter, resulting in $R\sim3000$ spectra. We chose a position angle (PA) of 90$^\circ$ based on telescope observing constraints and the position of the radio source in the galaxy. 

The CCDs on GMOS-N have two $\sim2\farcs8$--wide chip gaps, which correspond to $\sim100$~\AA\ gaps in the IFU spectra. We therefore took four 900~s observations, with two at each central wavelength setting of 580~nm and 585~nm. These two settings were chosen to capture all the lines of interest, avoid any emission lines falling within the chip gaps, and to fill out the continuum within the chip gaps.

Given the lack of point sources within the FOV of the science frames, we cannot assess the seeing on a frame-to-frame basis. However, all four frames used in the data reduction meet the following criteria: 85\% image quality, 50\% cloud cover, 80\% background and `any' water vapour. In practice, this means the seeing is likely within 1\arcsec.

In addition to the science frames, two 300~s standard star observations of Hz 44 were taken within the same observation window, using two different central wavelength settings: 580~nm and 600~nm. These two settings correspond to the lowest and highest central wavelength settings for all 10 objects in our Gemini observing program. We then use these two standard star observations to create a master sensitivity function that can be applied to all of our science observations. These frames are processed in a manner similar to the science data as described in Section~\ref{ssec:redux}.

\subsection{Basic Data Reduction}
\label{ssec:redux}
The data were reduced according to the GMOS IFU-1 data reduction tutorial\footnote{\url{https://gmos-ifu-1-data-reduction-tutorial-gemini-iraf.readthedocs.io/en/latest/}} \citep{gmosiraf,gmosifu}, which uses commands from the \texttt{gemini} package in pyraf \citep{pyraf}. We give a brief description of the process below.

Before reducing the science frames, we reduce the standard star observations to create the sensitivity functions. We first create a master bias for the standard star. We then correct one of two flats for bias and overscan, and use it to verify the mask definition file (MDF), which maps the fibers in the observation to their correct bundle, and masks out dead fibers. After defining the MDF, we trace the light illumination across the detector using the same flat. 

We use that reduced flat to extract the arc and calculate the wavelength solution, which is used to fully reduce the flat. This process includes modelling and subtracting scattered light between the bundles, correcting for quantum efficiency and calculating the response function.

After correcting the flat, we begin processing the standard star observation itself. We remove the bias and overscan, subtract scattered light and attach the MDF. The standard frames are then corrected for quantum efficiency, rectified to enforce the same wavelength scale, and then sky-subtracted. After this is completed, the reduced standard star frame is used to calculate the sensitivity function. This process is repeated for both standard star observations. The two individual sensitivity functions are combined to make the master sensitivity function, which can be applied to all science frames.

Finally, the science frame goes through the same processing as the standard star, with the addition of cosmic ray rejection using L.A.-cosmic \citep{vanDokkum2001}. We then use the sensitivity function calculated from the standard star observation to calibrate our science data. 

\subsection{Combining Exposures into Final Data Cube}

The basic data reduction process is completed for each individual exposure, creating 4 data cubes. Each data cube has a plate scale of 0\farcs0807 pixel$^{-1}$ in the two spatial directions, and $\sim0.5$~\AA\ pixel$^{-1}$ in the spectral direction. Due to the use of two different wavelength settings as discussed in Section~\ref{ssec:obs}, the two pairs of observations differ in their wavelength ranges by $\sim50$~\AA. In order to combine the exposures into the final data cube, we rebin all four individual cubes, using a flux-conserving method, to have the same wavelength scale and range. The final data cube has an observed wavelength range of $4133$--7232~\AA, with 0.5~\AA\ pixel$^{-1}$. By matching the wavelength range, we truncate the blue and red ends of the spectra with central wavelengths of 580~nm and 585~nm, respectively. However, this step does not impact any emission lines of interest.

The \texttt{IRAF}\footnote{IRAF is distributed by the National Optical Astronomy Observatories,
which are operated by the Association of Universities for Research in
Astronomy, Inc., under cooperative agreement with the National Science
Foundation.} task \texttt{imcombine} is then used to median combine the aligned images \citep{iraf}. The error bar for each pixel in the final IFU data cube is the standard deviation of the four values of that pixel, one from each exposure.

\subsection{Calculating WCS and Astrometry Correction}
The world coordinate system (WCS) for a GMOS IFU data cube is presented as physical coordinates relative to the central pixel. As precise astrometry is vital to properly identify the optical emission associated with the radio source, we performed an astrometry correction on our GMOS observations in a two step process.

In the first step, we convert the WCS from the physical system provided by the data reduction pipeline to an FK5 coordinate system using the information in question 3 of the data reduction section of the GMOS IFU question page\footnote{\url{http://ast.noao.edu/sites/default/files/GMOS_FAQ/GMOSIFU.html}}. The process is relatively straight-forward given that our PA is 90$^\circ$. The only deviation in our calculation from the process described on the GMOS IFU question page is that the GMOS detector plate scale has been updated from $0\farcs3$ pixel$^{-1}$ to $0\farcs0807$ pixel$^{-1}$ to match our observations. 

After constructing the initial WCS, we confirm the astrometry by comparing the 2--D image of the GMOS observation to the $r$-band image of the galaxy from SDSS. The $r$-band image is chosen to match the region of the spectrum where the GMOS signal is strongest, i.e. $\lambda\gtrsim6000$~\AA. While the DECaLS observation is deeper than SDSS, the GMOS observation is close to the nucleus of the galaxy, which is well-detected in both DECaLS and SDSS.

We collapse the GMOS IFU cube in the spectral direction, creating a 2--D spatial profile. The peaks of the 2--D GMOS profile and the $r$-band photometry of SDSS are aligned to perform the final astrometry correction. Given the larger pixel size of SDSS, the astrometry is correct to within 0\farcs2.
\section{2--D Spatial Emission}
\label{sec:2dem}
\begin{figure*}
    \centering
    \includegraphics[width=0.48\textwidth]{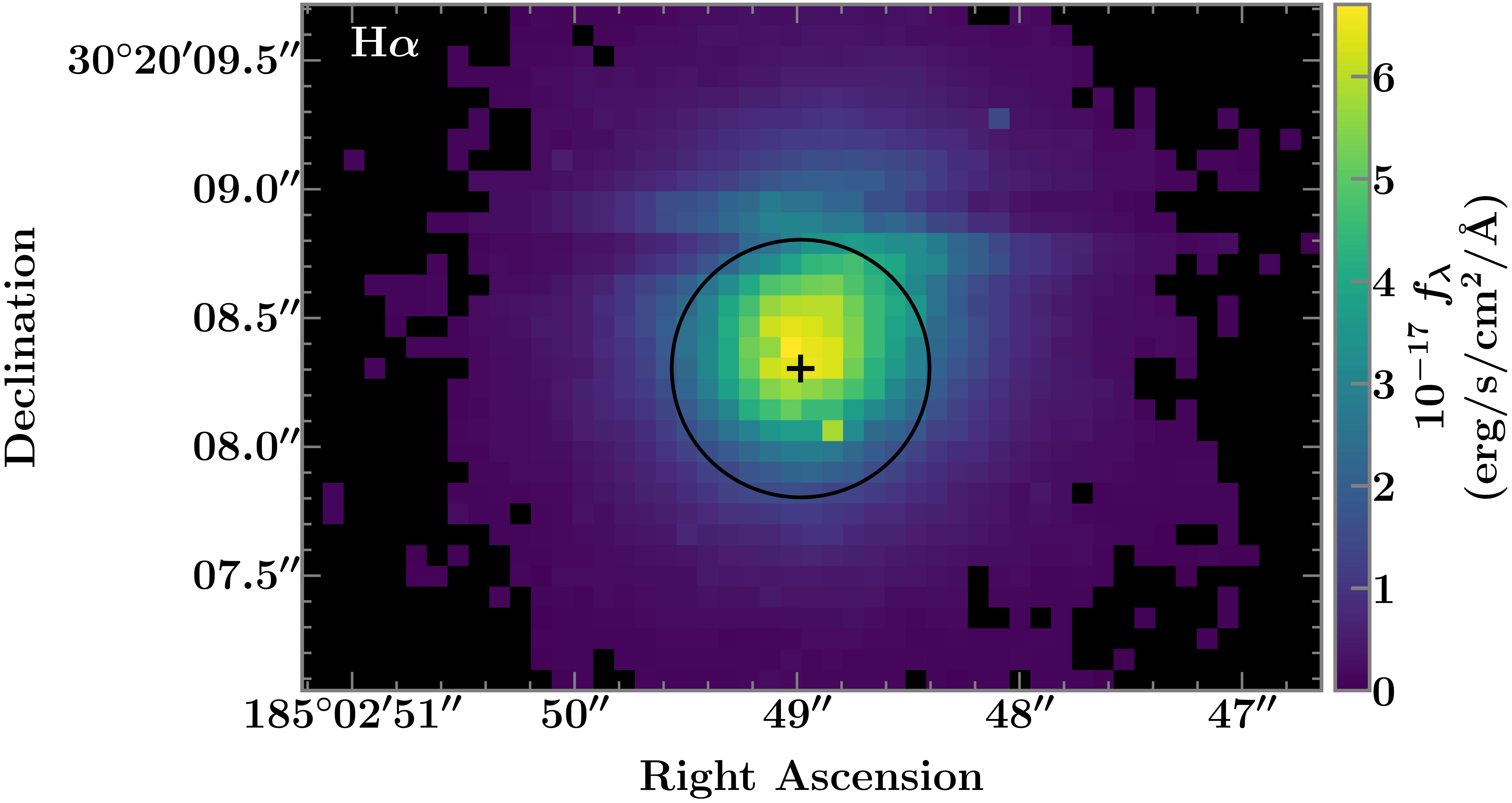}
    \includegraphics[width=0.48\textwidth]{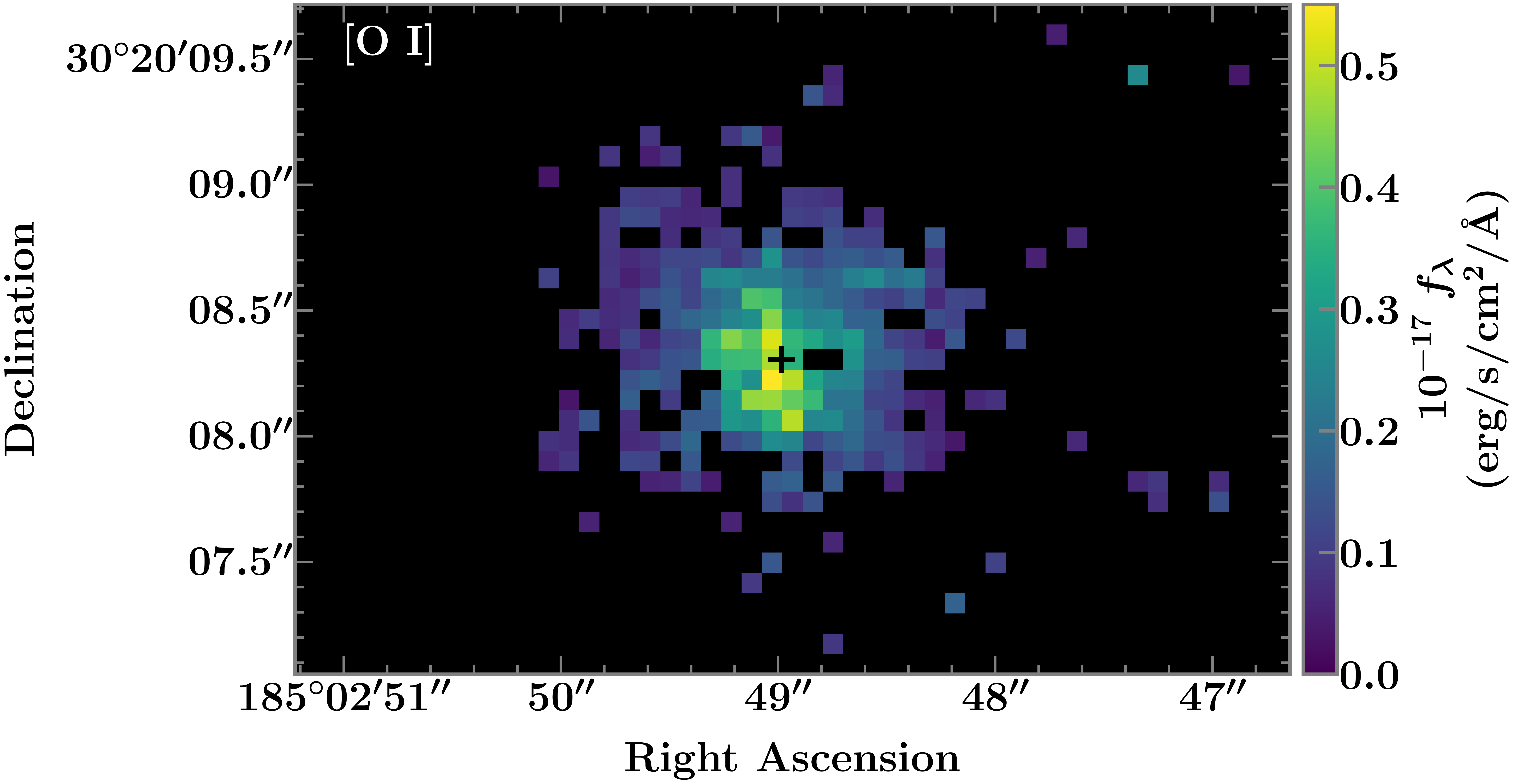}
        \includegraphics[width=0.48\textwidth]{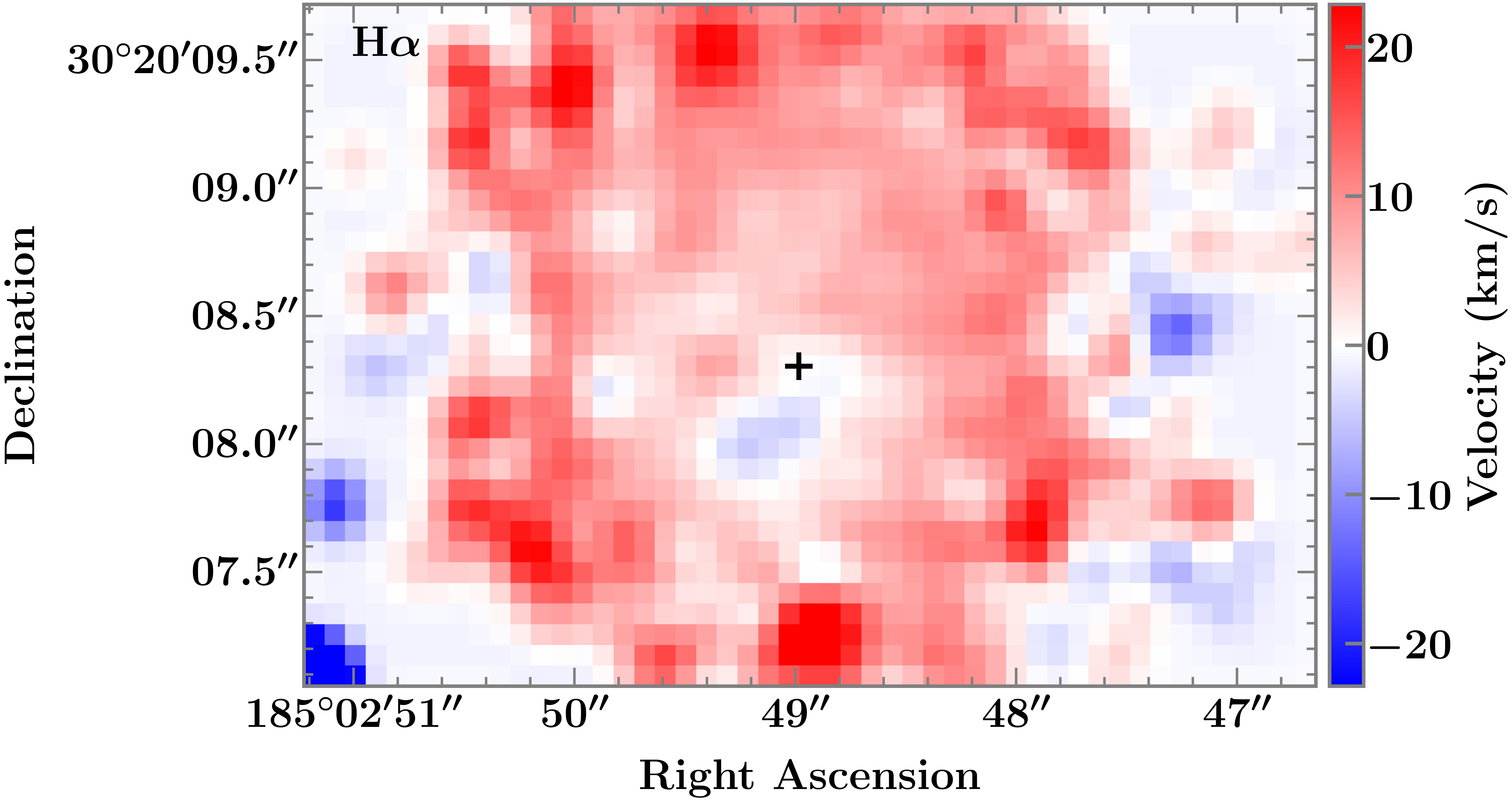}
    \includegraphics[width=0.48\textwidth]{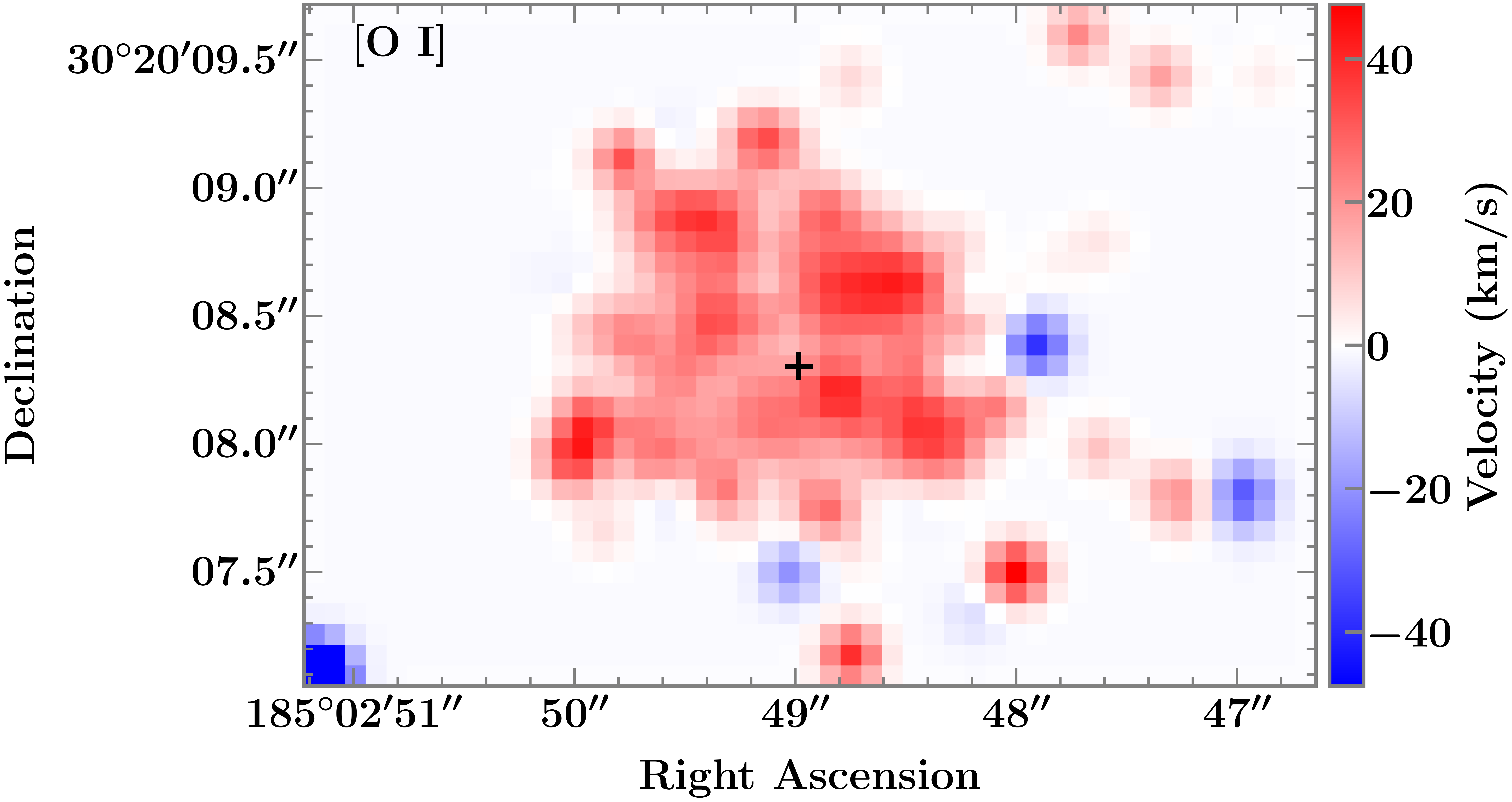}
    \caption{\textit{Top: }2--D Flux maps for H$\alpha$ (left) and [\ion{O}{1}]$\lambda$ 6300 (right) for J1220+3020. The flux maps are presented in units of $10^{-17} f_\lambda$, with the color bar on the right of the image. A black pixel represents spectra where the line was not detected. The shifted rows are due to a small mismatch of fibers in the data reduction, but this effect does not significantly impact our 1-D resolved measurements. The flux was calculated via fitting a combination of Gaussians, as described in Section~\ref{sec:2dem}. We show the shape of the point spread function via the black aperture in the H$\alpha$ flux panel. The black cross in each panel marks the position of the radio source. The H$\alpha$ emission peaks near the radio source, and there is an elongated feature of enhanced [\ion{O}{1}] emission centralized on the radio source. This [\ion{O}{1}] feature could be indicative of an outflow. \textit{Bottom: }Same as Top except for the velocity. The velocity is measured in km~s$^{-1}$, and presented relative to that at the radio source. We have smoothed the velocity maps with a 2--D Gaussian, where $\sigma_x=\sigma_y=1$ spatial pixel. The radio source appears to be the origin for a blue-shifted knot of H$\alpha$ gas, but this offset is consistent with the measured errors in velocity ($\sigma_v\approx10$~km~s$^{-1}$). Meanwhile there is no clear trend in the [\ion{O}{1}] emission.}
    \label{fig:2d_maps}
\end{figure*}

Here we examine the {observed} 2--D flux and velocity maps for both H$\alpha$ and [\ion{O}{1}]$\lambda$6300, as presented in Figure~\ref{fig:2d_maps}. The velocity maps are smoothed using a 2--D Gaussian function, where $\sigma_x=\sigma_y=1$ spatial pixel. 
While [\ion{O}{3}]$\lambda$5007 is typically used to study gas kinematics, the noise at the blue end of our spectra make the measurements of [\ion{O}{3}] using one spatial pixel unreliable. Additionally, we do not include velocity dispersion maps as most of the line widths are comparable to the instrumental resolution. Finally, we do not correct for reddening because we are interested in qualitative global trends. Since H$\alpha$ and [\ion{O}{1}] are relatively close to each other, any changes due to reddening will be approximately the same for both lines.

We calculated these maps by fitting a combination of Gaussians to both the H$\alpha$ and [\ion{O}{1}]$\lambda$6300 emission line for all 1--D spectra in the IFU cube. We simultaneously fit the Gaussians with a second order polynomial to account for the underlying continuum. As some of the spectra required more than one Gaussian, we followed the fitting routine described below.

For each spectrum, we initially fit the line with one Gaussian. If the signal-to-noise (S/N) of the integrated flux is at least 3 for H$\alpha$ or 2 for [\ion{O}{1}], then the line is considered detected and a second fit using two Gaussian is performed. We accept the 2-Gaussian fit if the reduced $\chi^2$ is lower by at least 20\%. For the pixels with spectra that displayed blending between the [\ion{N}{2}] doublet and H$\alpha$, all three lines were fit simultaneously. In some cases, a third Gaussian was needed to fully describe the emission from H$\alpha$. When more than one Gaussian was needed, the centroid is defined as the peak of the observed emission line. 

The results of this process are shown in Figure~\ref{fig:2d_maps}, with the position of the radio source shown as the black cross. The velocity maps are calculated with respect to the centroid of the emission line at the position of the radio source. We note a small mismatch in the fibers, which can be clearly seen in the H$\alpha$ 2--D map. This effect does not significantly impact our 1--D aperture measurements, which are the basis of our quantitative analysis. The radio source is well-aligned with the peak in the H$\alpha$ flux along with a knot of blue-shifted H$\alpha$ gas. However, its velocity offset is approximately equal to the error in the velocity maps, ($\sigma_v\approx10$~km~s$^{-1}$). Additionally, the radio source appears to be in the center of an elongated feature of enhanced [\ion{O}{1}] emission, which could indicate an outflow. Unlike H$\alpha$ we do not see any trends in the velocity of the [\ion{O}{1}] gas. However, we do see broadened [\ion{O}{1}] emission associated with the elongated, enhanced feature, as discussed in Section~\ref{ssec:o1fex}.

We note that while these 2--D maps are useful for examining the spatial extent of emission lines within the galaxies, the individual pixels are not resolved. Therefore, we avoid creating 2--D emission line maps of different components in these emission lines as they may not be physically meaningful. 

\section{Resolved 1--D Spectra}
\label{sec:1dres}
\subsection{Region definitions}\label{ssec:regdef}
\begin{figure*}
    \centering
    \includegraphics[width=0.48\textwidth]{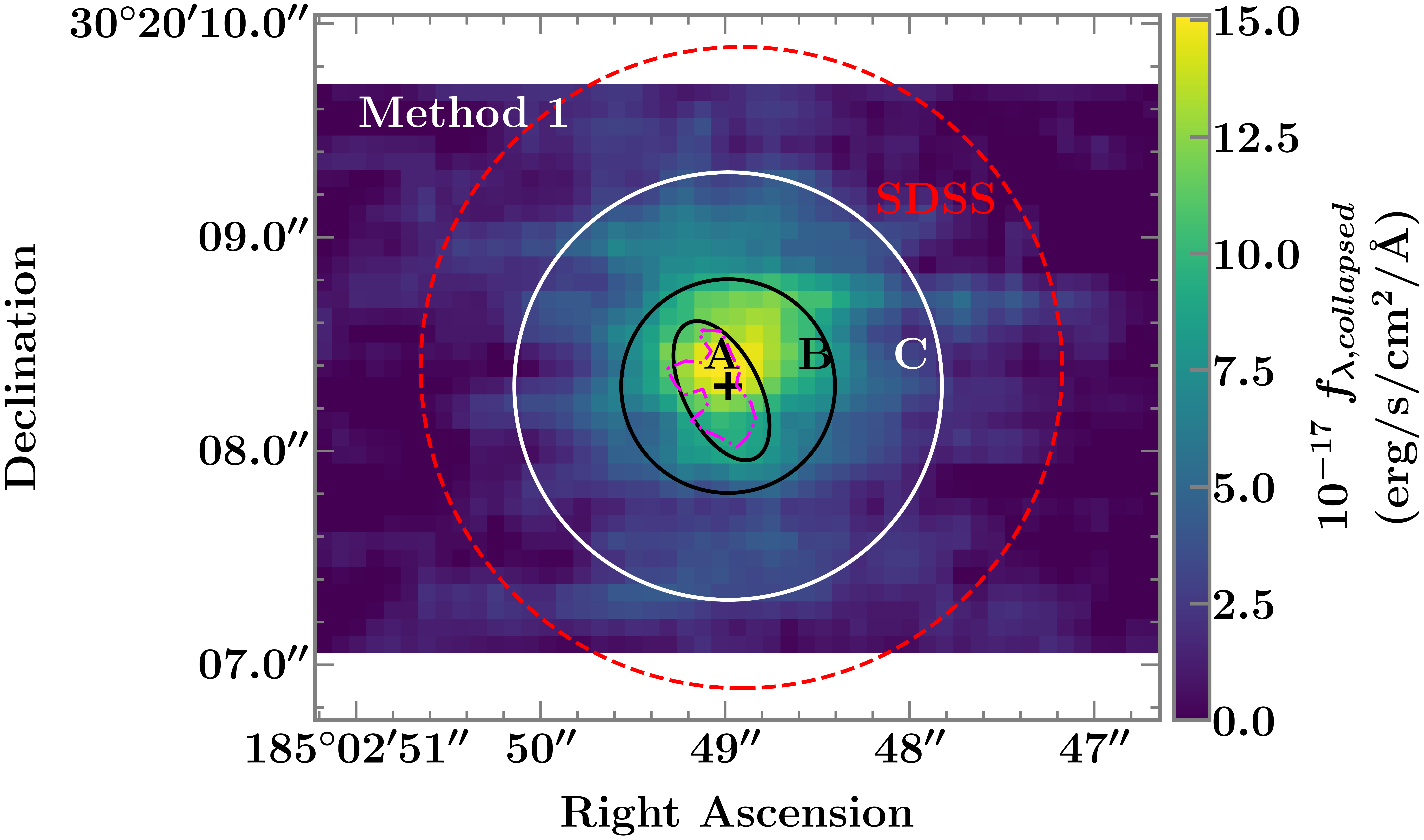}
    \includegraphics[width=0.48\textwidth]{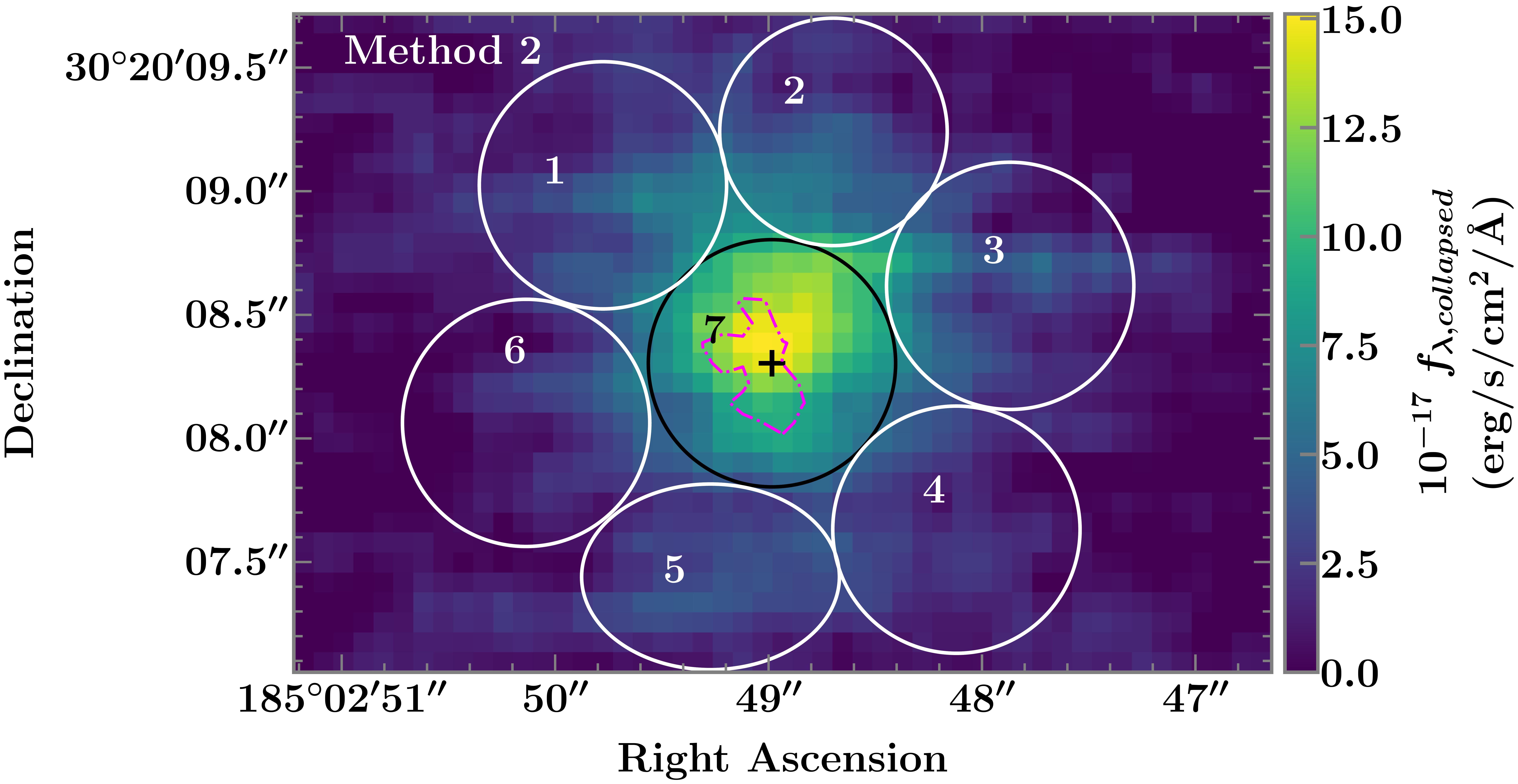}
\caption{The collapsed IFU image of J1220+3020, showing the different 1--D regions used in the analysis of this work. The black cross is the radio source, and the magenta dot-dash contour shows the position of the enhanced [\ion{O}{1}] emission within the galaxy, as seen in Figure~\ref{fig:2d_maps}. The definitions of this contour and its major axis are given in Section~\ref{ssec:regdef}. All the regions are designated by a letter or a number, and will be referenced in the analysis in Section~\ref{sec:discussion}. \textit{Left: }First aperture definition method, with 3 total apertures that all overlap. The first is the elliptical aperture around the enhanced [\ion{O}{1}] feature. Then there are apertures with 1\arcsec and 2\arcsec diameters centered on the radio source. The aperture of the SDSS spectrum is shown as a red dashed line for reference. \textit{Right: }Second aperture definition method. Aperture 7 is identical to aperture B in the first method, and is a 1\arcsec aperture around the radio source. Apertures 1 and 4 follow the direction of the major axis of the enhanced [\ion{O}{1}] feature, while the remaining apertures fill in the space between these two regions.}
    \label{fig:coll_image}
\end{figure*}

In order to accurately measure the emission from the radio source, we must spatially bin the data to create resolved 1--D spectra. We binned the data using two different methods: the first focuses on the integrated emission, while the second explores spatially resolved regions. The positions of these regions within the IFU FOV in are shown Figure~\ref{fig:coll_image}, and the aperture definitions are provided in Table~\ref{table:regions}.

The first method, shown in the left panel, consists of concentric, cumulative apertures. Aperture A isolates the region of enhanced [\ion{O}{1}] emission seen in Figure~\ref{fig:2d_maps}. We used the matplotlib contour command \citep{matplotlib} to quantify the extent of the enhanced [\ion{O}{1}] emission, shown as the magenta contour, which is includes all pixels with flux $f_i$ where $f_{i}\ge 0.65f_{\rm max}$. We confirmed by eye that the contour encompassed the enhanced [\ion{O}{1}] emission feature seen in Figure~\ref{fig:2d_maps}. We then define Aperture A to match the general shape of the [\ion{O}{1}] feature, and use the major axis of that aperture to define the major axis of the [\ion{O}{1}] emission.

Meanwhile, apertures B and C are cumulative circular apertures of increasing radius centered on the radio source. The largest aperture shown is the measured SDSS DR8 spectrum. Each aperture in method 1 includes all of the light within the defined area. If there is an AGN at the position of the radio source, these apertures will illustrate whether its light can be hidden by increasing the contamination from the host galaxy. 

The second method is shown on the right panel, and involves resolved regions around the radio source. Aperture 7 is the same as aperture B in method 1. We used the major axis of aperture A from method 1 to define apertures 1 and 4 in method 2. In other words, apertures 1 and 4 measure the emission that is directly outside of the radio source aperture, and in the direction of the potential [\ion{O}{1}] outflow. We then filled in the gaps between regions 1 and 4 using apertures that were spatially resolved and outside the radio source aperture.

\begin{figure*}
    \centering
    \includegraphics[width=0.98\textwidth]{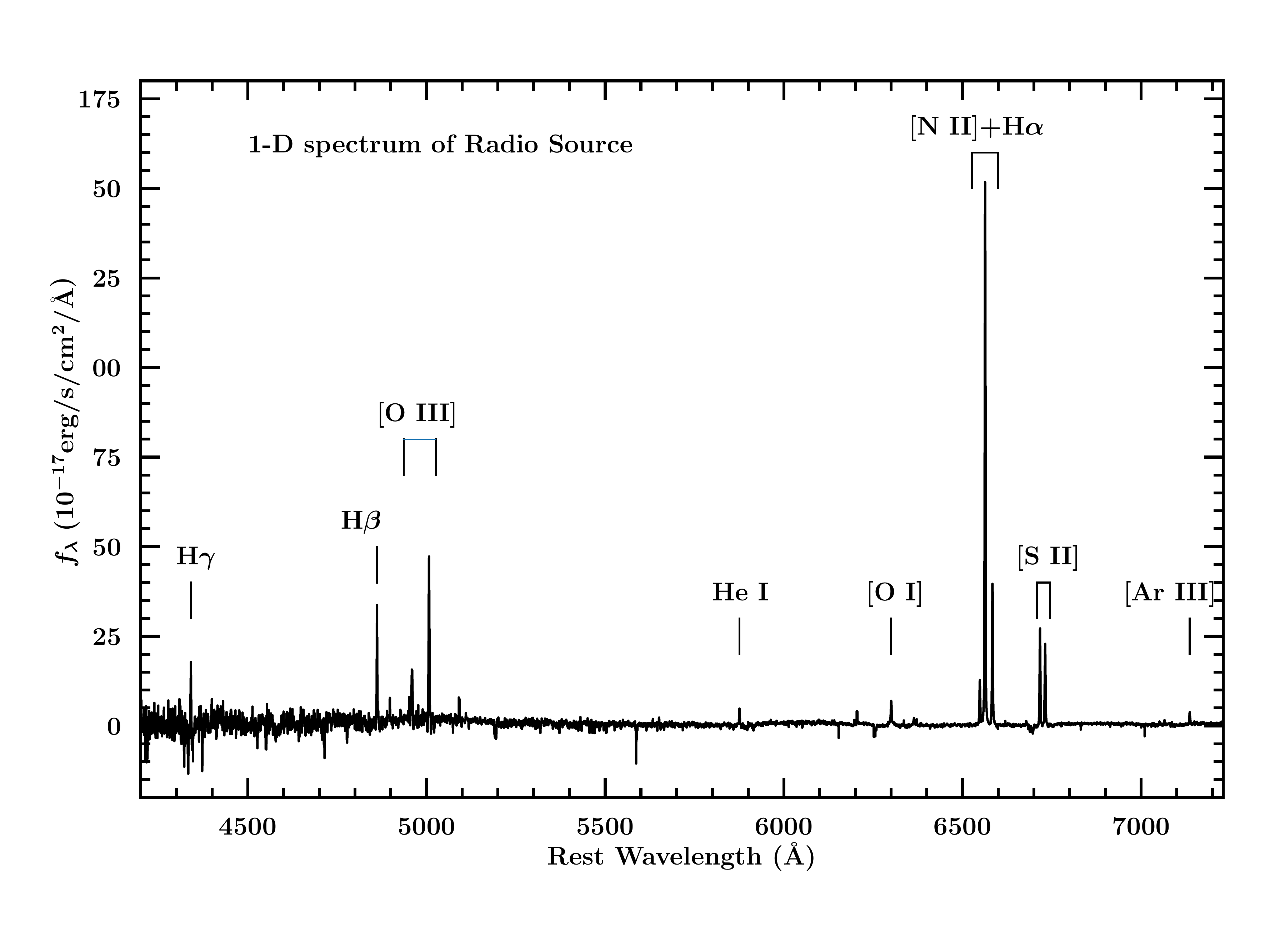}\vspace{-1.cm}
    \caption{The resolved continuum-subtracted, 1\arcsec spectrum of the radio source, or aperture B in method 1 and aperture 7 in method 2, with the strong lines labeled. The spectrum is a combination of three different CCD chips, with the chip gaps at $\sim5300$\AA\ and $\sim6000$\AA. The first CCD chip, covering the blue end of the spectrum, is systematically noisier than the other two. However, all of the traditional lines used for narrow-line diagnostic diagrams are clearly detected, and there is strong [\ion{O}{1}]$\lambda$6300 emission, which is indicative of non-stellar processes.}
    \label{fig:1dspec}
\end{figure*}

We note that the resolution of the GMOS IFU data is 1\arcsec, while the radio source is a point source of size $0\farcs18\times0\farcs17$. Therefore, all apertures except for aperture A in method 1 are resolved, and cover a much larger physical area than the radio source. We show the resolved 1\arcsec spectrum of the BH candidate (aperture B in method 1 and aperture 7 in method 2) in Figure~\ref{fig:1dspec}.

\begin{deluxetable*}{ccccccccl}
  \tablecaption{Aperture Definitions for 1--D Spectra\label{table:regions}}
\tabletypesize{\footnotesize}
\setlength{\tabcolsep}{4pt}
\renewcommand{\arraystretch}{1.}
\tablewidth{0pt}
\tablehead{
\colhead{Method\tablenotemark{a}} &{Number} & {R.A.} & {Dec.} & {a\tablenotemark{b}} & {b} &{$\theta$} &{Area} & {Description}\vspace{-1mm}\\
{} & {} & {(deg)} & {(deg)} & {(arcsec)} & {(arcsec)} & {(deg)} & {(kpc$^2$)} &{}\vspace{-1mm}}
\startdata
{1} & {A} & {185.04695} & {30.33563} & {0.35} & {0.18} &{90} & {0.059} & {Captures Enhanced [\ion{O}{1}] emission}\\
{1} & {B} & {185.04694} & {30.33564} & {0.50} & {...} &{...} & {0.234} & {1\arcsec Aperture centered on Radio Source}\\
{1} & {C} & {185.04694} & {30.33564} & {1.00} & {...} &{...} & {0.936} & {2\arcsec Aperture centered on Radio Source}\\
{2} & {1} & {185.04716} & {30.33584} & {0.50} & {...} & {...} & {0.234} & {Follows semi-major axis of enhanced [\ion{O}{1}] emission}\\
{2} & {2} & {185.04686} & {30.33590} & {0.46} & {...} & {...} & {0.198} & {Off-set from radio source in NNW direction}\\
{2} & {3} & {185.04663} & {30.33573} & {0.50} & {...} & {...} & {0.234} & {Off-set from radio source in NW direction}\\
{2} & {4} & {185.04670} & {30.33545} & {0.50} & {...} & {...} & {0.234} & {Follows semi-major axis of enhanced [\ion{O}{1}] emission}\\
{2} & {5} & {185.04702} & {30.33540} & {0.52} & {0.38} & {0} & {0.253} & {Off-set from radio source in SSE direction}\\
{2} & {6} & {185.04726} & {30.33557} & {0.50} & {...} & {...} & {0.234} & {Off-set from radio source in SE direction}\\
{2} & {7\tablenotemark{c}} & {185.04694} & {30.33564} & {0.50} & {...} &{...} &{0.234} &  {1\arcsec Aperture around Radio Source}\\
\enddata
\tablenotetext{a}{The apertures in Method 1 are cumulative, and have total areas of 0.197 arcsec$^2$,0.78~arcsec$^2$, and 3.14~arcsec$^2$, respectively.}\vspace{-2mm}
\tablenotetext{b}{If The aperture is circular, ``a'' represents the radius. If it is an ellipse, then ``a'' is the semi-major axis.}\vspace{-2mm}
\tablenotetext{c}{Aperture 7 in method 2 is identical to aperture B in method 1.}
\end{deluxetable*}

\subsection{Continuum subtraction and emission-line measurements of 1--D spectra}\label{ssec:elines}
After extracting the 1--D spectra, we subtract the continuum and measure the emission lines; the narrow line flux measurements used in our analysis are reported in Table~\ref{table:fluxes}. As there is no clear detection of the stellar continuum, we fit a third-order polynomial to the continuum for all spectra. We then fit groups of lines individually with the code with \texttt{pyspeckit} \citep{pyspeckit}, which is a wrapper around the python package MPFIT, and is designed for fitting astronomical spectra. We employ a fitting process similar to that of \citet{reines2013}. We show the fits to all of the strong lines and a table of the line component widths for aperture B, i.e., the spectrum of the radio source, in Appendix~\ref{app:apb_info}.

\begin{figure}
    \centering
    \includegraphics[width=0.48\textwidth]{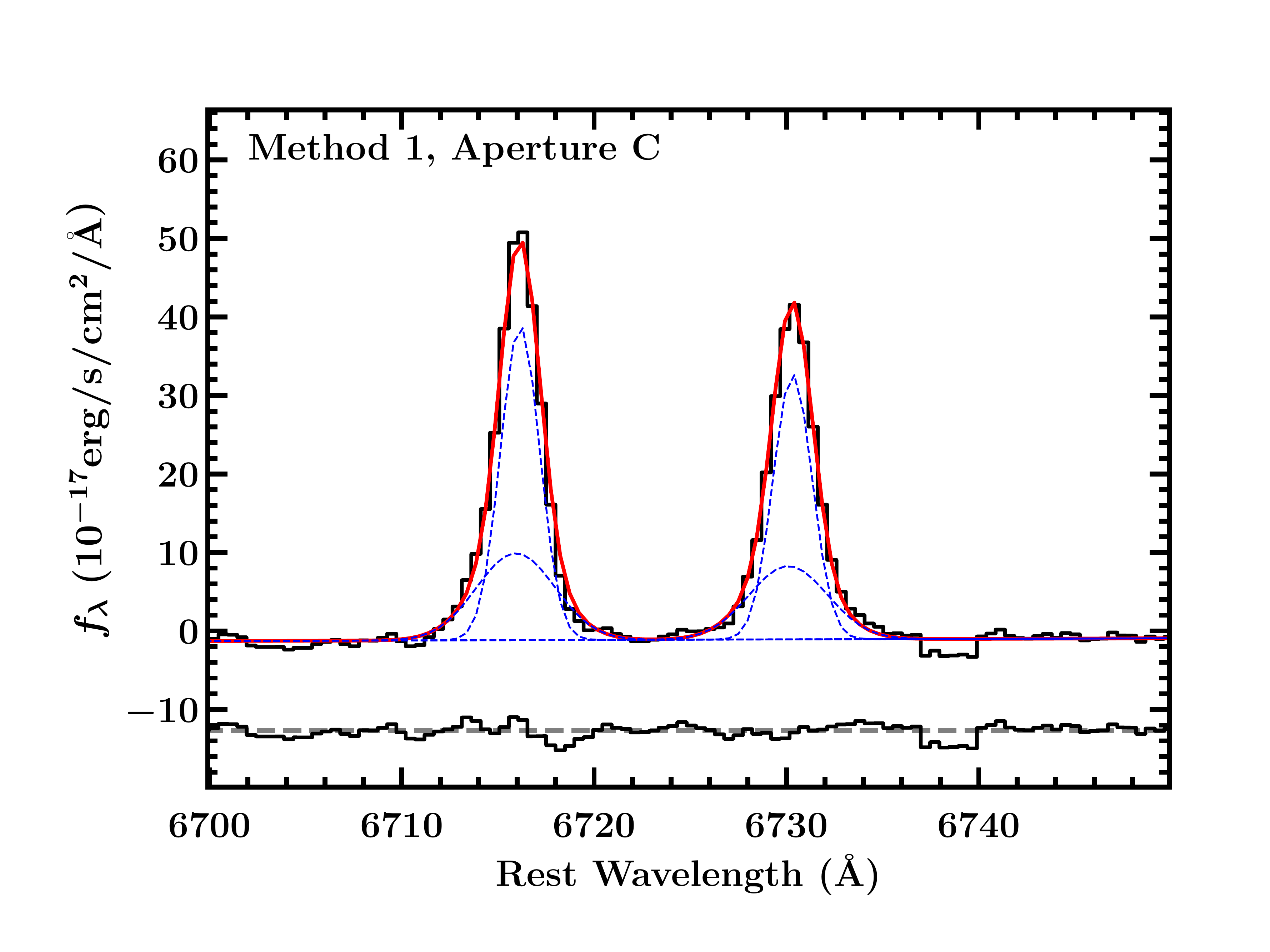}
    \includegraphics[width=0.48\textwidth]{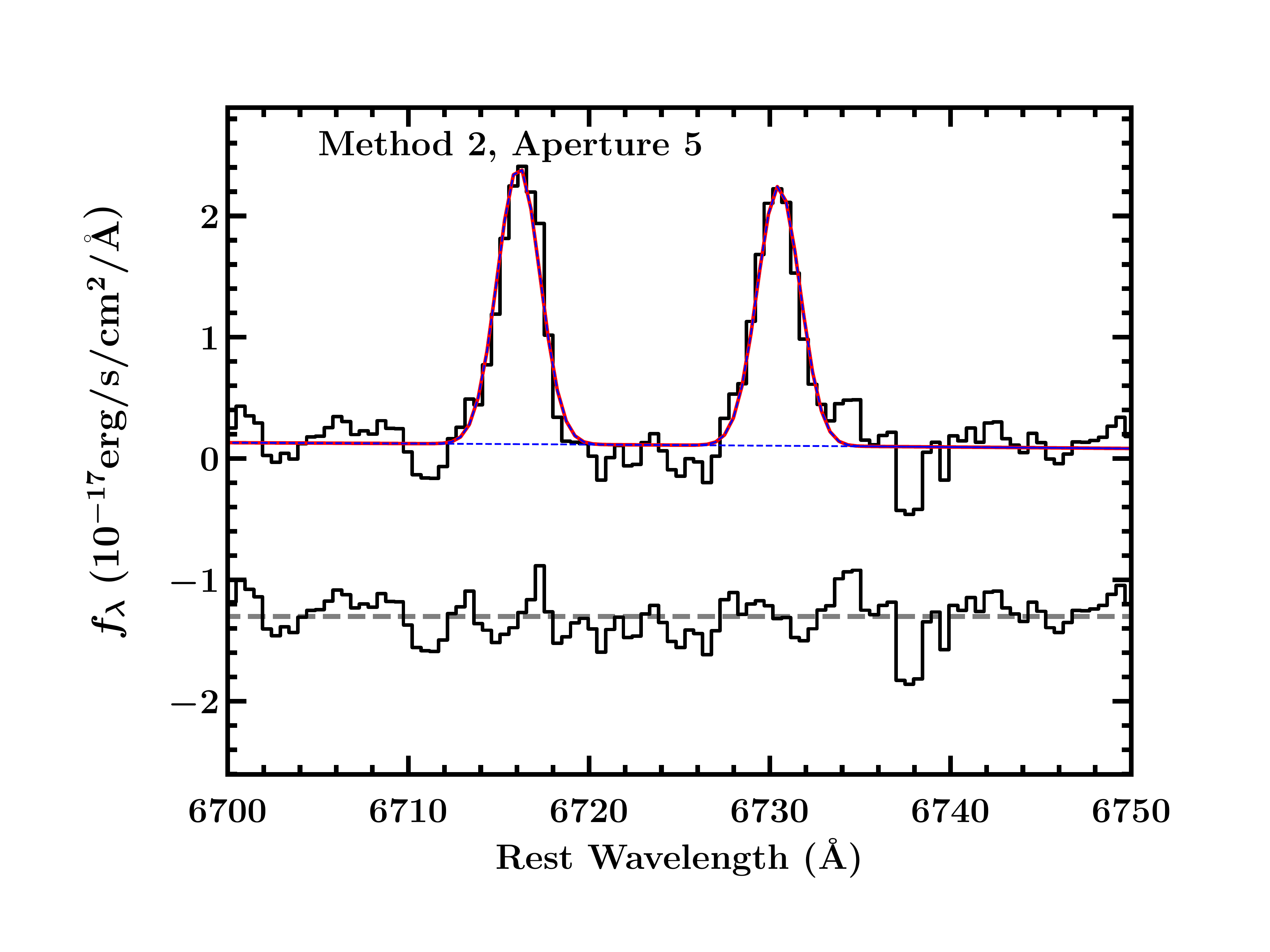}
    \caption{\textit{Top: }Example of two-component fit to the [\ion{S}{2}] lines, using the aperture C from method 1 (2\arcsec aperture). The data are in black, and the residuals are shown below the fit. The red line is the final fit to the emission lines, while the blue dashed lines show the individual components and the baseline fit. This spectrum includes emission from the radio source. \textit{Bottom: }Same as the top, but for a one-component fit to the [\ion{S}{2}] lines, using the fifth aperture from method 2. This spectrum is not coincident with the radio source.}
    \label{fig:s2_fits}
\end{figure}

We begin by fitting the [\ion{S}{2}]$\lambda\lambda$ 6716,6731 doublet. We employ a linear fit to the continuum to correct any issues with the original continuum fit, and two Gaussians, one for each line in the doublet. The Gaussians are constrained to have the same width, and a separation defined by the their laboratory wavelengths. The [\ion{S}{2}] doublet is then fit with a two-component Gaussian model, where the width, relative position, and height ratio for the two-component model constrained to be the same for each line. We select the two-component model if the reduced $\chi^2$ value is at least 10\% lower than that of the single Gaussian model. We found that the only spectra that required a two-component model for the [\ion{S}{2}] doublet were those that included emission from the radio source. This includes the unresolved aperture A and resolved apertures B and C in method 1. Examples of one and two-component Gaussian fits for [\ion{S}{2}] are shown in Figure~\ref{fig:s2_fits}.

After choosing the appropriate the [\ion{S}{2}] model, we use it to constrain the [\ion{N}{2}]+H$\alpha$ complex. We again employ a linear fit to the continuum and use the [\ion{S}{2}] model as a template for the narrow line emission in both the [\ion{N}{2}] doublet and H$\alpha$. For the [\ion{N}{2}] doublet, we require the emission lines to have the same width as the [\ion{S}{2}] model in velocity space, constrain the separation as defined by their laboratory wavelengths, and fix the relative flux as [\ion{N}{2}]$\lambda6584/$[\ion{N}{2}]$\lambda6548=3$. H$\alpha$ is a bit harder to constrain via [\ion{S}{2}] as it is a recombination line, not a forbidden line. If [\ion{S}{2}] is best-fit by a single Gaussian model, we use it to constrain the narrow-line emission of H$\alpha$, but allow the width of H$\alpha$ to increase by as much as 25\%. If [\ion{S}{2}] is best-fit by a two-component Gaussian model, we constrain the H$\alpha$ narrow emission to have the same profile and width (in velocity space) as [\ion{S}{2}]. In some cases, another broader component was needed to fit the H$\alpha$ emission line. We accepted the fit with the broad component if the reduced $\chi^2$ was smaller by at least 20\%. Since H$\alpha$ is a recombination line, we did not assume this third component was necessarily a ``broad'' component from the broad line region of an AGN. A more complete discussion of this issue is presented in Section~\ref{ssec:broadlines}.

We then apply the same process used for the [\ion{N}{2}] doublet to the [\ion{O}{1}] doublet, with the assumed flux ratio of [\ion{O}{1}]$\lambda6300/$[\ion{O}{1}]$\lambda6363=3$. However, we found that for the spectra that included emission from the radio source, a third component was needed to fully fit the [\ion{O}{1}] emission lines. Additionally, the [\ion{Fe}{10}]$\lambda$ 6374 emission line was detected near the radio source. Both of these issues are discussed in more detail in Section~\ref{ssec:broadlines}.

Finally, we fit the H$\beta$ and [\ion{O}{3}] doublets separately from the rest of the emission lines. These lines fall in the noisier part of the GMOS spectra. We therefore fit them independently, using no constrains on H$\beta$. The [\ion{O}{3}]$\lambda\lambda$ 4959,5007 doublet was constrained so that their widths are the same, their separation is defined by their laboratory wavelengths, and their flux ratio set as [\ion{O}{3}]$\lambda5007/$[\ion{O}{3}]$\lambda4959=3$. In some spectra, H$\beta$ was not detected. Due to the higher level of noise on the blue end of the spectrum, we accept fits to H$\beta$ down to a ${\rm S/N}=2$. Below that threshold, we calculate the 3-$\sigma$ upper limit, using the [\ion{O}{3}]$\lambda5007$ line as a template.
\subsection{Features of the Line-Emitting Gas}
\label{ssec:broadlines}
\subsubsection{Broad Hydrogen Balmer Emission}
Given that H$\alpha$ is a recombination line, it is typically emitted physically closer to the power source than [\ion{S}{2}]. As a result, H$\alpha$ will potentially have a broader profile, and will likely have a different emission-line shape than the forbidden narrow lines. As we are interested in using BPT diagrams in our analysis, which relies on the non-broad line region emission, we must be cautious when differentiating between the broad and narrow components. \citet{reines2013} accounted for this issue by allowing the width of H$\alpha$ to increase by as much as 25\% compared to that of the [\ion{S}{2}] doublet, but given the higher spectral resolution of our data, we allowed a broader component to be included if needed. We accepted that component if the reduced $\chi^2$ was smaller by at least 20\%. 

\begin{figure}
    \centering
    \includegraphics[width=0.48\textwidth]{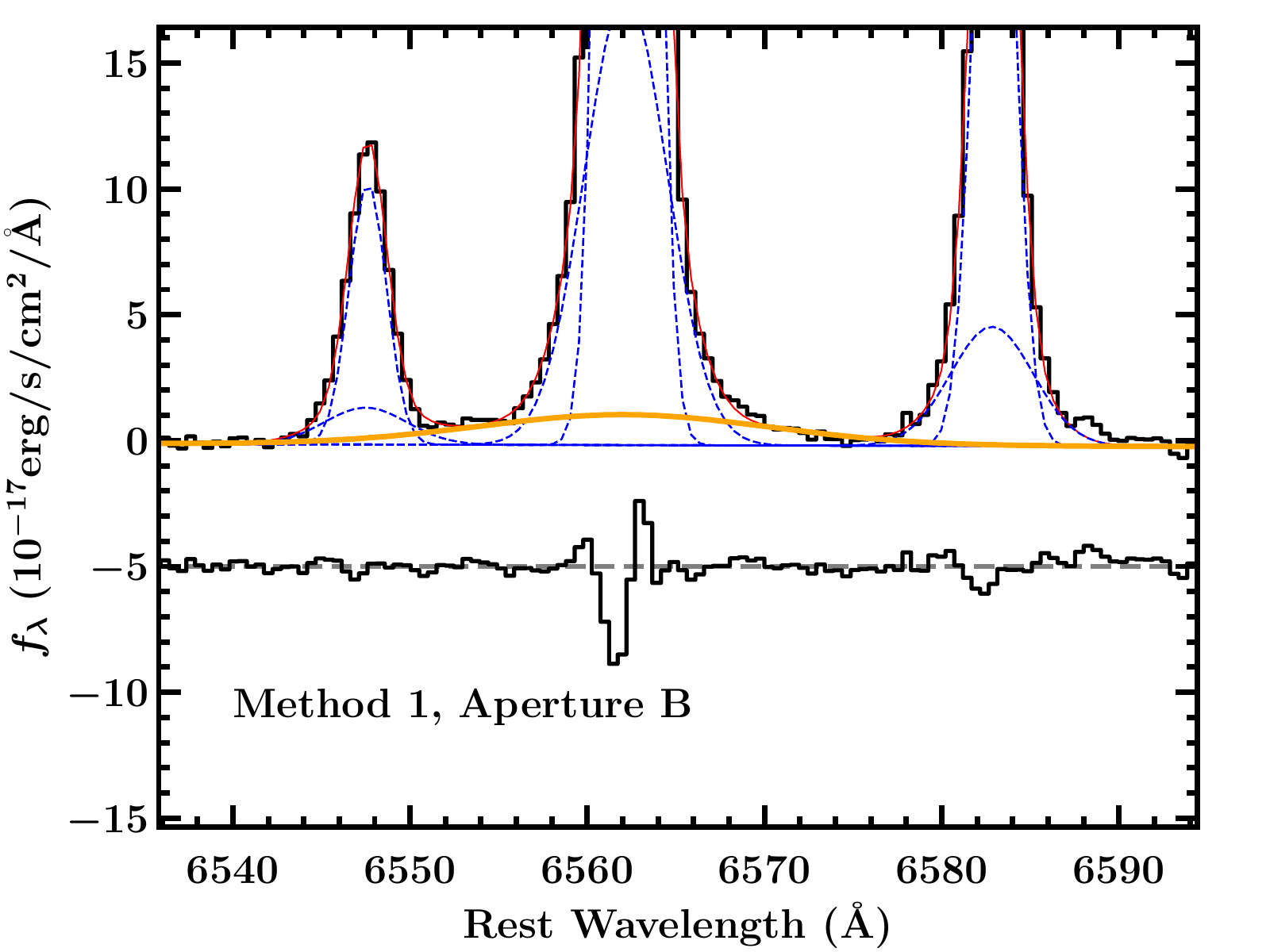}
    \includegraphics[width=0.48\textwidth]{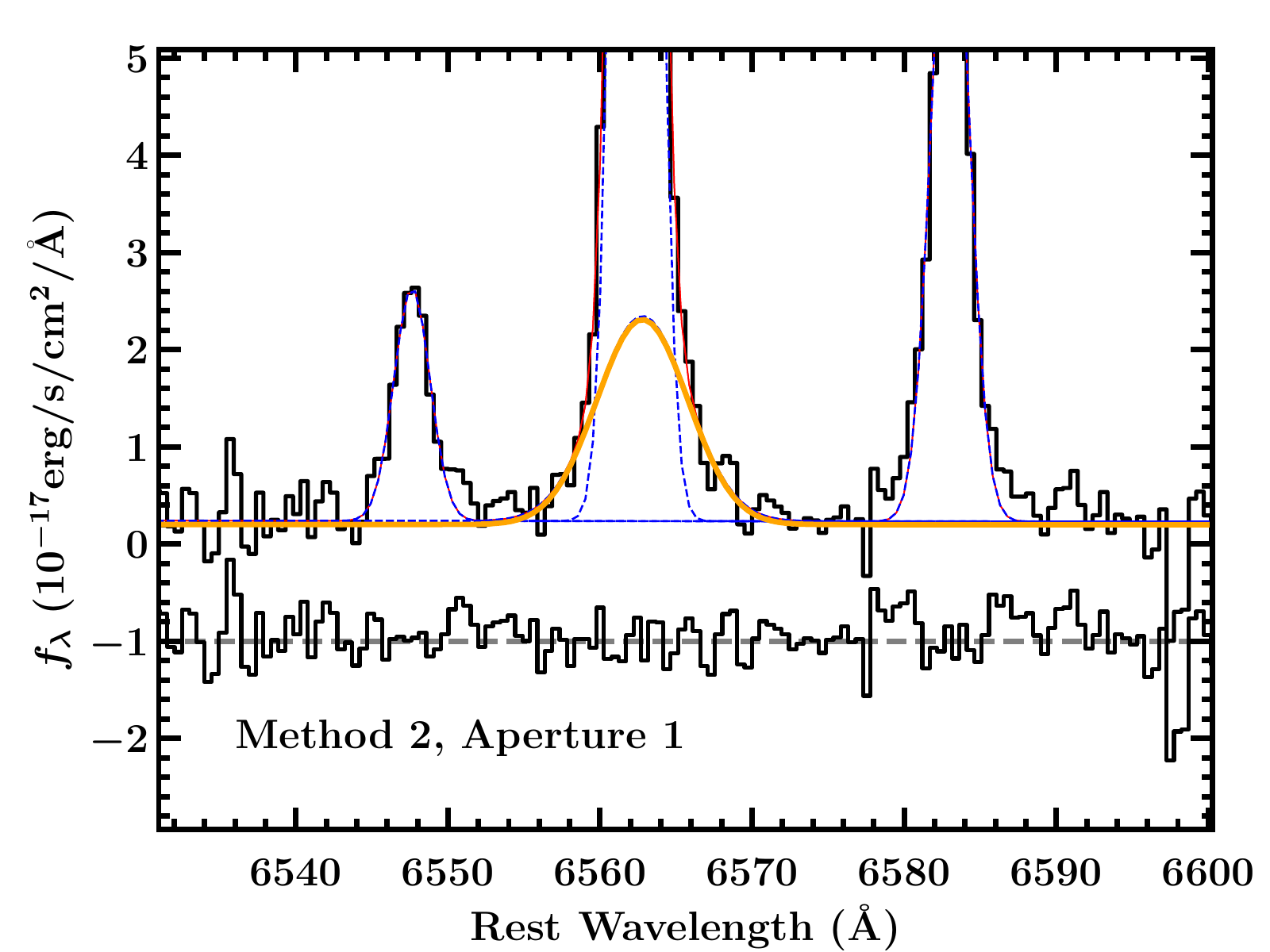}
    \caption{{\it Top: }The fit to the H$\alpha$ broad component for method 1, aperture B, the 1\arcsec aperture around radio source. The data are shown in black, the full model is shown in red, the narrow line components are shown by the blue dashed lines and the H$\alpha$ broad component is the solid orange line. The residuals are shown below the fit. This broad component has a ${\rm FWHM}\sim900$~km~s$^{-1}$ after correcting for instrumental broadening. We consider this Gaussian to be broad-line emission, and do not include it in the narrow-line measurement. {\it Bottom: }Same as top, but for method 2, aperture 1, a 1\arcsec aperture offset from the radio source. The ``broad'' component has a ${\rm FWHM}\sim300$~km~s$^{-1}$ after correcting for instrumental broadening. We therefore do {\it not} consider this Gaussian to be broad-line emission, and we {\it do} include it in the narrow-line measurement.}
    \label{fig:habroad}
\end{figure}

The only two spectra in method 1 that required an extra component for H$\alpha$ were apertures A (unresolved aperture around enhanced [\ion{O}{1}] feature) and B (resolved, 1\arcsec aperture of radio source). In both cases, the full width at half maximum, or FWHM $\gtrsim900$~km~s$^{-1}$ which are clearly broad-line components. We also measure the H$\alpha$ flux in a 1\arcsec annulus directly outside the radio source, and do not detect this broad component. We therefore conclude that this broad emission is associated with the radio source. The fit for aperture B is shown in the top panel of Figure~\ref{fig:habroad}. We note that the H$\alpha$ broad component in aperture B (which is the same as aperture 7) is only 5\% of the total observed H$\alpha$ emission, and therefore removing this feature will not significantly affect the emission line ratios used in our analysis.

Conversely for method 2, all apertures required an extra component to fully describe the H$\alpha$ emission. However, only aperture 7 (which is the same as aperture B in method 1) had a FWHM greater than 500~km~s$^{-1}$, before correcting for instrumental broadening. We therefore assume that this extra H$\alpha$ component in apertures 1--6 are narrow emission, and include them in our analysis. We show an example of the secondary H$\alpha$ component for aperture 1 in the bottom panel of Figure~\ref{fig:habroad}. 

We find no evidence for a broad component in H$\beta$. However, if the broad component is contributing to the observed H$\beta$ emission, it could cause an overestimation of the narrow-line flux. If we assume that the broad H$\beta$ component contributes the same proportional amount of flux as that in H$\alpha$ ($\sim5$\%), then any overestimation of the H$\beta$ flux falls within the calculated flux uncertainty making this effect negligible.

\subsubsection{[\ion{O}{1}] and [\ion{Fe}{10}] emission}\label{ssec:o1fex}
In addition to the broad H$\alpha$ component seen near the radio source, we also detect both a broader [\ion{O}{1}]$\lambda$~6300 component and [\ion{Fe}{10}]$\lambda$~6374. [\ion{Fe}{10}]$\lambda$~6374 is a known AGN coronal line \citep{penston1984,Netzer2013}, with an ionization potential of 262.1~eV \citep{Oetken1977}. Furthermore, the ratio of [\ion{Fe}{10}]$\lambda$~6374/H$\beta$ can be used to calculate the velocity of jet-driven shocks in AGNs \citep{wilson1999}. The integrated flux of the [\ion{Fe}{10}] coronal line is detected with a ${\rm S/N = 4.3}$, and the peak of the line is $\approx7\sigma$ above the surrounding continuum.

\begin{figure}
    \centering
    \includegraphics[width=0.49\textwidth]{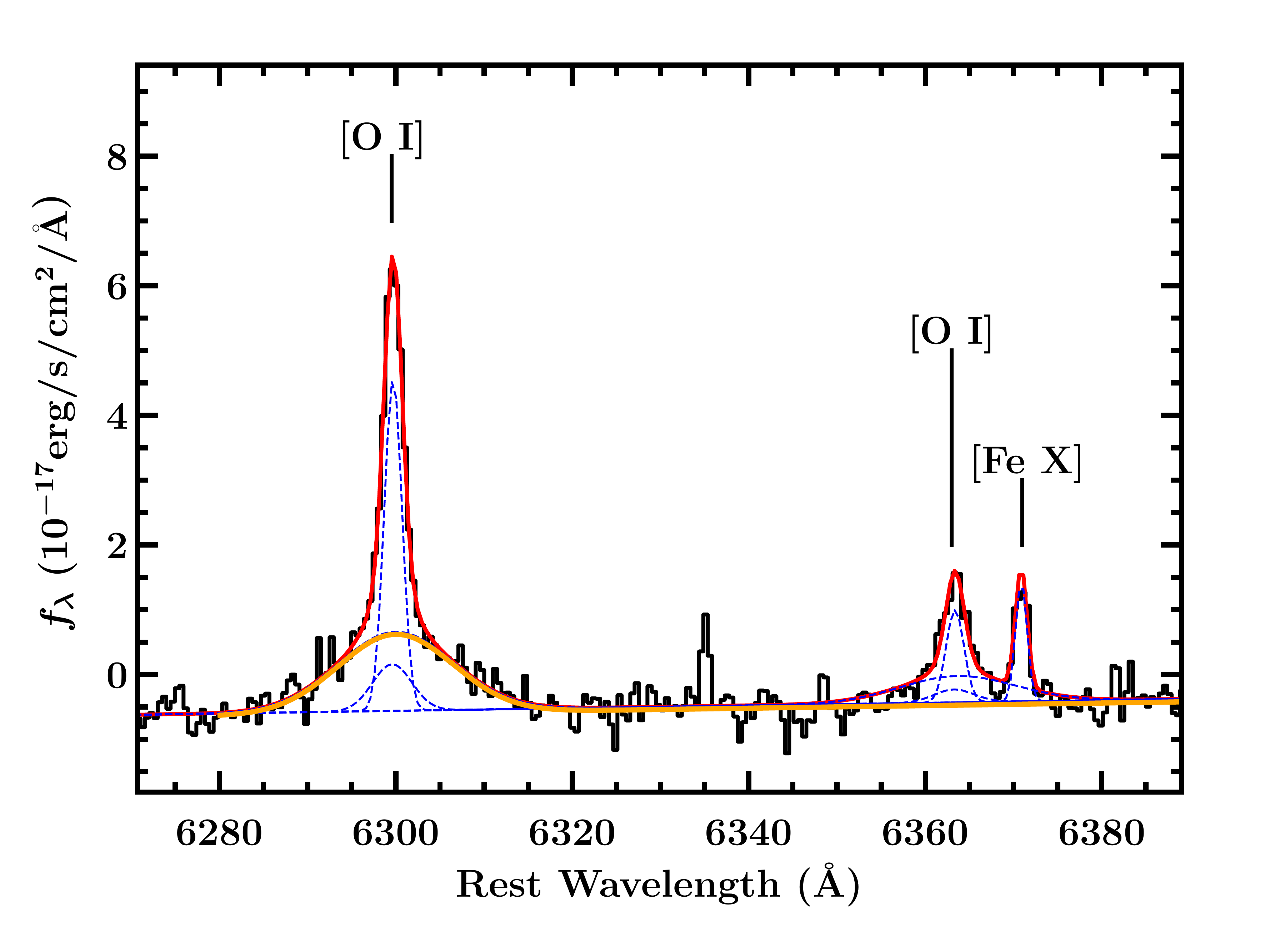}
    \caption{Same as Figure~\ref{fig:habroad}, but for the [\ion{O}{1}] broad emission from method 1, aperture B. We label both the [\ion{O}{1}]$\lambda\lambda$6300,6363 doublet and the [\ion{Fe}{10}]$\lambda$6374 emission line for reference. The [\ion{O}{1}] broad component, shown in orange, cannot be explained by the [\ion{S}{2}] profile template alone. The width of this component is ${\rm FWHM}\sim700$~km~s$^{-1}$, after correcting for instrumental broadening. Since the forbidden line emission cannot originate in the broad line region, we do not exclude it from our analysis. [\ion{Fe}{10}]$\lambda$6374 is a known AGN coronal line; the integrated flux measurement has a ${\rm S/N}=4.3$ and the peak of the line is $\approx7\sigma$ above the continuum.}
    \label{fig:o1broad}
\end{figure}

Meanwhile, LINERs with LLAGNs and strong radio emission are known to have enhanced [\ion{O}{1}] emission which can have different profiles from that of [\ion{S}{2}] \citep{Balmaverde2014}. We note that we only detect the broader [\ion{O}{1}] and [\ion{Fe}{10}] emission in apertures coincident with the radio source. While we do not see similar kinematic components in the other strong lines, [\ion{O}{1}] is not as easily created by stellar sources than the other forbidden lines which makes it less subject to galaxy contamination. An example fit to the [\ion{O}{1}]+[\ion{Fe}{10}] complex is shown for method 1, aperture B in Figure~\ref{fig:o1broad}. While the additional [\ion{O}{1}] component is broader than those seen in the other forbidden lines, forbidden line emission cannot originate in the broad line region. We therefore do not exclude them from our narrow-line analysis.

\subsubsection{Overall physical conditions of the gas}
After measuring the emission lines, we calculated the electron densities of all 1--D spectra using the [\ion{S}{2}]$\lambda\lambda$~6716,6731, assuming an electron temperature $T_e=10^4$~K. All of the spectra had densities within the range of $n_e=500$--$1000$~cm$^{-3}$, and showed no clear spatial trend. We additionally measured the typical velocity widths of the narrow forbidden lines to be $\textrm{FWHM}\sim100$--300~km~s$^{-1}$. As the spectral resolution is $\sim100$~km~s$^{-1}$, the narrow lines are marginally resolved. Finally, the metallicity was calculated using equation 1 in \citet{pettini2004}. We found the metallicity to be $12+\log(O/H)=8.55$--8.61, which is sub-solar. We will use these physical properties to constrain the shock models in Section~\ref{ssec:shock_setup}.

\section{Comparison of Spectra to Models}
\label{sec:mod_comp}
\subsection{Standard BPT Diagnostics}
In order to determine the mechanism powering the observed emission, we compared the measured emission lines to commonly used models in the standard BPT diagrams. As these line ratios are reddening insensitive, we do not correct for dust.

\begin{figure*}
    \centering
    \includegraphics[width=0.98\textwidth]{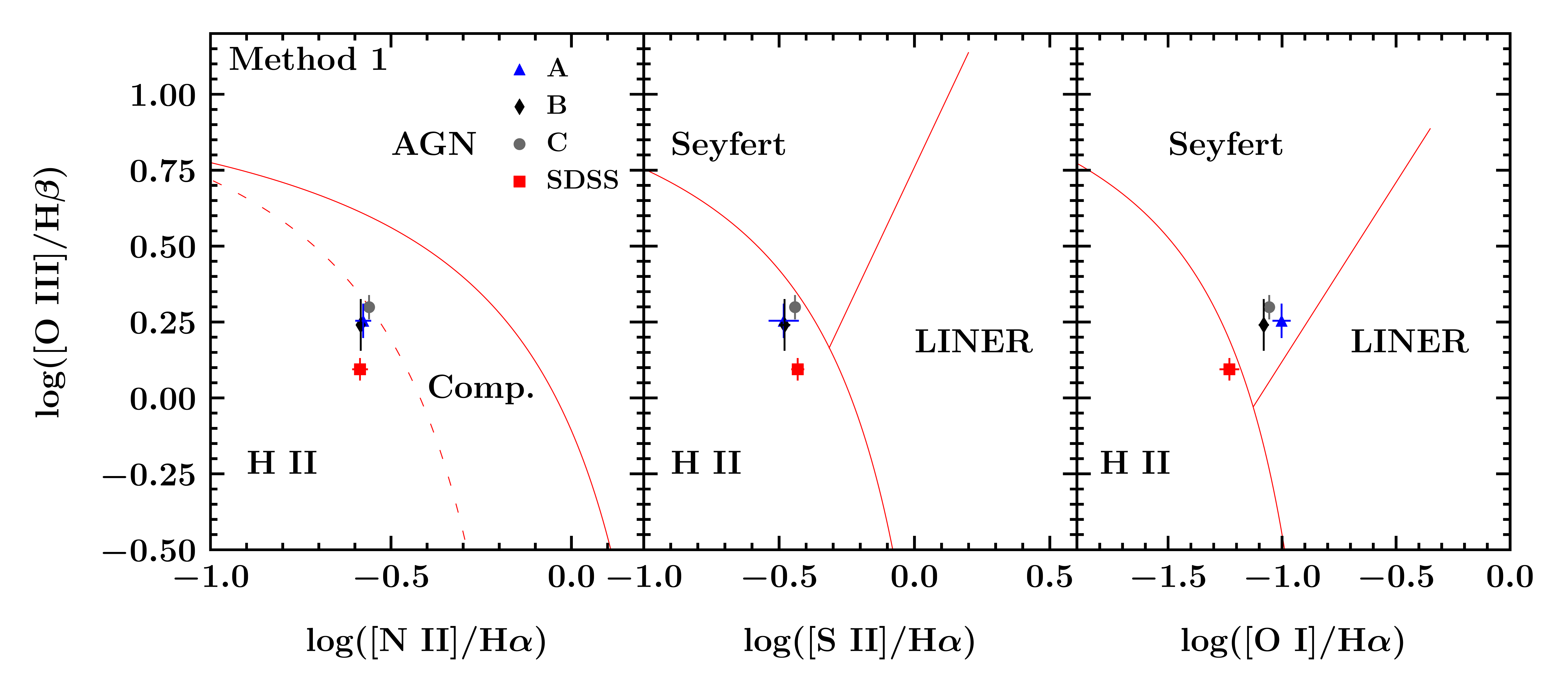}
    \includegraphics[width=0.98\textwidth]{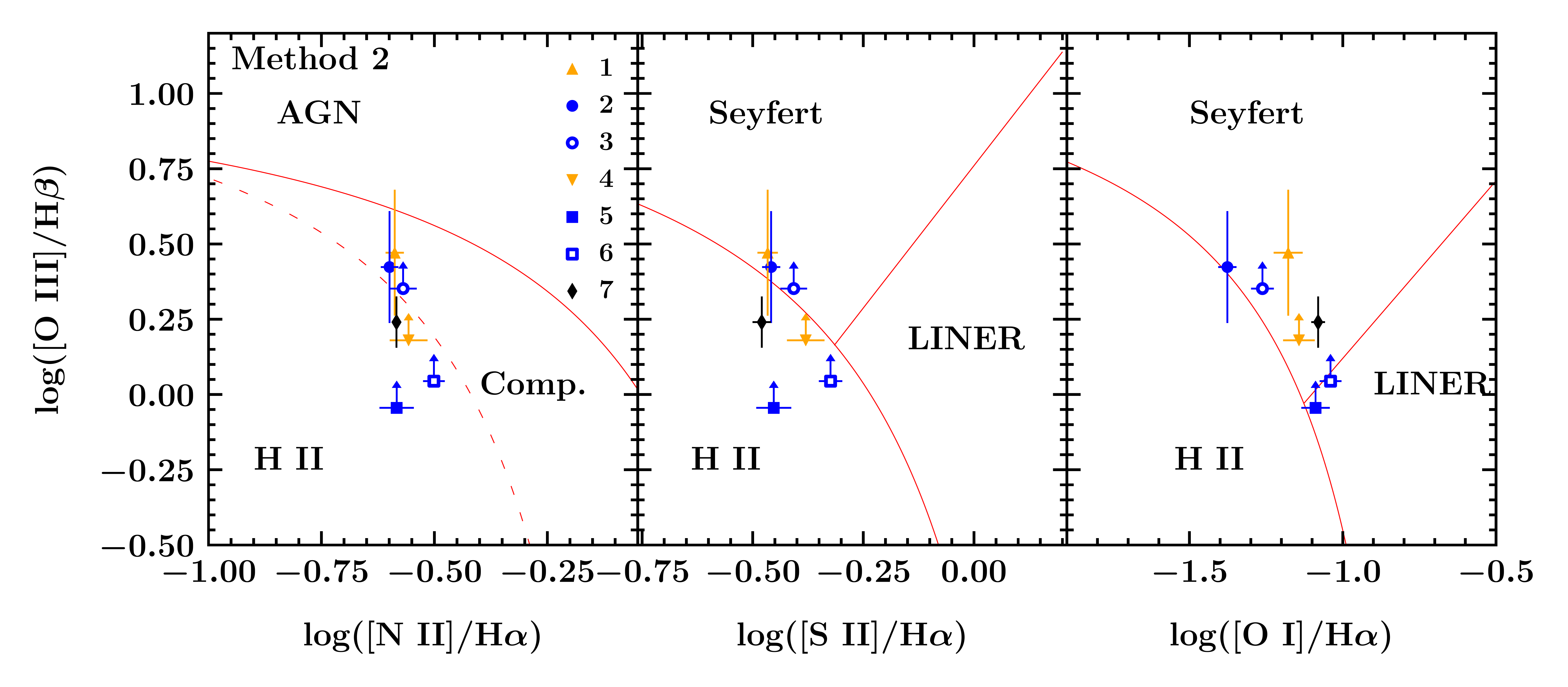}
    \caption{{\it Top: }The [\ion{O}{3}]/H$\beta$ vs.~[\ion{N}{2}]/H$\alpha$, [\ion{S}{2}]/H$\alpha$ and [\ion{O}{1}]/H$\alpha$ diagrams, or BPT diagrams, for the apertures from method 1. The markers representing the different spectra are defined by the legend in the [\ion{N}{2}]/H$\alpha$ diagram. The SDSS spectrum is from \cite{reines2020}. The red solid lines are the extreme starburst lines and the Seyfert lines as defined in \cite{kewley2006}. The dashed line in the [\ion{N}{2}]/H$\alpha$ diagram is the composite line from \cite{Kauffmann2003}. The labels indicate the different classification regions for each diagram: \ion{H}{2}, composite and AGN for [\ion{N}{2}]/H$\alpha$, and \ion{H}{2}, Seyfert and LINER for [\ion{S}{2}]/H$\alpha$ and [\ion{O}{1}]/H$\alpha$. There is clearly enhanced [\ion{O}{1}]/H$\alpha$ emission that cannot be described by star formation. Therefore, there is no clear mechanism among all three diagrams. {\it Bottom: }Same as top, but for method 2. We again see enhanced [\ion{O}{1}]/H$\alpha$ emission not explained by star formation, creating ambiguity in these diagnostics.}
    \label{fig:bpt}
\end{figure*}

We compare the [\ion{N}{2}]/H$\alpha$, [\ion{S}{2}]/H$\alpha$ and [\ion{O}{1}]/H$\alpha$ vs.~[\ion{O}{3}]/H$\beta$ measurements for all apertures for both methods to the diagnostics for gas photoionized by hot stars in Figure~\ref{fig:bpt}. The \citet{kewley2006} theoretical extreme starburst lines are shown in solid red, while the \citet{Kauffmann2003} empirical composite line for the [\ion{N}{2}]/H$\alpha$ diagram is shown as a dashed red line. We present the results for each method below.

\begin{itemize}
    \item[]{\it Method 1 --} The apertures in this method represent the ``integrated'' measurements, with aperture A as the smallest region and aperture C as the largest. The SDSS measurements from \citet{reines2020} are also plotted for reference. 
    
    From Figure \ref{fig:bpt}, we see the [\ion{N}{2}]/H$\alpha$ and [\ion{S}{2}]/H$\alpha$ ratios are consistent with ionization by hot stars. However, all of the [\ion{O}{1}]/H$\alpha$ measurements from the GMOS data presented in this work fall inside the Seyfert locus. This is in contrast to the SDSS observation that has a smaller exposure time and worse spectral resolution ($R\sim2000$ for SDSS vs.~$R\sim3000$ for GMOS), and so it is not surprising that a weak AGN would be difficult to detect. While there is ambiguity in these results when comparing to the three standard BPT diagrams, 
    [\ion{O}{1}]/H$\alpha$ is the cleanest diagnostic to discriminate between stellar and non-stellar processes since it is sensitive to the hardness of the ionizing radiation field \citep{kewley2006}. 
    \item[]{\it Method 2 --} This method includes the resolved 1\arcsec radio source spectrum (aperture 7), and resolved, off-nuclear spectra (apertures 1-6). The two regions that are directly outside the endpoints of the enhanced [\ion{O}{1}] feature are shown as orange arrows (apertures 1 and 4), while the regions surrounding the radio source are shown in blue (apertures 2, 3, 5 and 6). The black diamond (aperture 7) is the same as aperture B in method 1. We note that apertures 1 and 4 do not show systematically different line ratios from apertures 2, 3, 5 and 6. This could imply that the outflow does not reach the region outside the 1\arcsec central aperture.
    
    The [\ion{O}{1}]/H$\alpha$ emission is again enhanced in this diagram, with almost all points consistent with Seyfert-like line ratios. Similarly, the [\ion{N}{2}]/H$\alpha$ and [\ion{S}{2}]/H$\alpha$ are largely consistent with the hot stars and composite loci. As in method 1, there is no single mechanism that can describe all of the observed emission.
\end{itemize}

\subsection{Shocked Emission Diagnostics}
\label{ssec:shock_setup}
The spectra in J1220+3020 do not clearly produce AGN-like emission as defined by the three standard BPT diagrams (with the exception of the [\ion{O}{1}]/H$\alpha$ diagram). However, there is a well-detected broad H$\alpha$ component and [\ion{Fe}{10}] line that are clear indicators of an AGN, creating ambiguity in these diagnostics. As other LLAGNs have been well-described by shock models \citep[see][and references therein]{Kewley2019}, we also compared our data to the shock and shock+precursor models from \citet{allen2008}. We describe the models and parameter constraints below, and compare the data to the models in Section~\ref{sssec:shockmod}.

The shock models were produced using the Mappings III code \citep{allen2008,sutherland2013} and consist of two physical scenarios: simple shocks and shocks with a precursor. For the simple shocks model, the gas is collisionally ionized by the propagating shock. Meanwhile, in models with a precursor, the shock-heated gas is allowed to produce photons that travel upstream and ionize the gas ahead of the shock front. The shock with a precursor model is usually invoked for situations where both photoionization and shock excitation are present \citep[e.g., in LINERs;][]{Molina2018}.

The Mappings III models cover a wide range of magnetic field strengths, velocities, densities and metallicities. While the radio source could have a magnetic field strength of $B\gtrsim50~\mu$G, this would only hold within the $<0\farcs2$ source itself. The resolved regions that we explore have a minimum size of 1\arcsec, which is much larger than the radio source, so the magnetic field strength could be very low. We therefore chose to set the magnetic field strength at $1~\mu$G, as its grid overlapped with most of the models incorporating magnetic fields of less than $10~\mu$G. In general, AGNs can have a wide range of radio strengths \citep[see Table 1 of][]{Ho2008}, and the subsequent magnetic field strength is dependent on both the strength and morphology of that radio emission \citep[see equation 5.109 in][]{condon2016}.

In order to set the velocity constraint, we examined the widths of the narrow emission lines. The forbidden lines typically had $\textrm{FWHMs}\sim100$--300~km~s$^{-1}$. However, the [\ion{O}{1}] line was slightly broader, especially in the radio source spectrum, which had a separate broad component with $\textrm{FWHM}\sim700$~km~s$^{-1}$. If shocks are indeed driving the observed emission, the observed low velocity widths of the forbidden lines imply that the shock should also have a low velocity. We therefore constrained the velocities in the models to be within the range $v=100$--500 km~s$^{-1}$, where the upper limit is the typical boundary between low- and high-velocity shocks.

The density and metallicity parameters are jointly explored in the \citet{allen2008} models. For all metallicities that are not solar, a density of $n=1$~cm$^{-3}$ is assumed. Only the solar metallicity models explore changes in density, from $n=0.1-10^3$~cm$^{-3}$. The galaxy studied here has a sub-solar metallicity, determined by the [\ion{N}{2}]/H$\alpha$ ratio, and a density of $n\approx10^3$~cm$^{-3}$, given by the [\ion{S}{2}]$\lambda\lambda$~6716,6731 doublet. Therefore, the models used here do not cover the region of parameter space occupied by our data.

To compensate for this inconsistency between density and metallicity, we have created a mixed model grid. We include models with solar abundances and densities of $n=1-1000$~cm$^{-3}$, as well as models with a fixed density of $n=1$~cm$^{3}$ that assume metallicities consistent with the Large and Small Magellanic Clouds (LMC and SMC). 

Finally, for the radio source spectrum, we detect the AGN coronal line [\ion{Fe}{10}]$\lambda$~6374. The [\ion{Fe}{10}]$\lambda$~6374/H$\beta$ ratio can be used to calculate the jet-driven shock velocities \citep{wilson1999}. After correcting for reddening, we find the [\ion{Fe}{10}]$\lambda$~6374/H$\beta\sim0.23$, which implies a shock velocity in the range $v=200$--300~km~s$^{-1}$, which is consistent with the measured velocities seen in the narrow forbidden lines. 

\subsubsection{Comparison to Shock Models}\label{sssec:shockmod} 
\begin{figure*}
    \centering
    \includegraphics[width=0.48\textwidth]{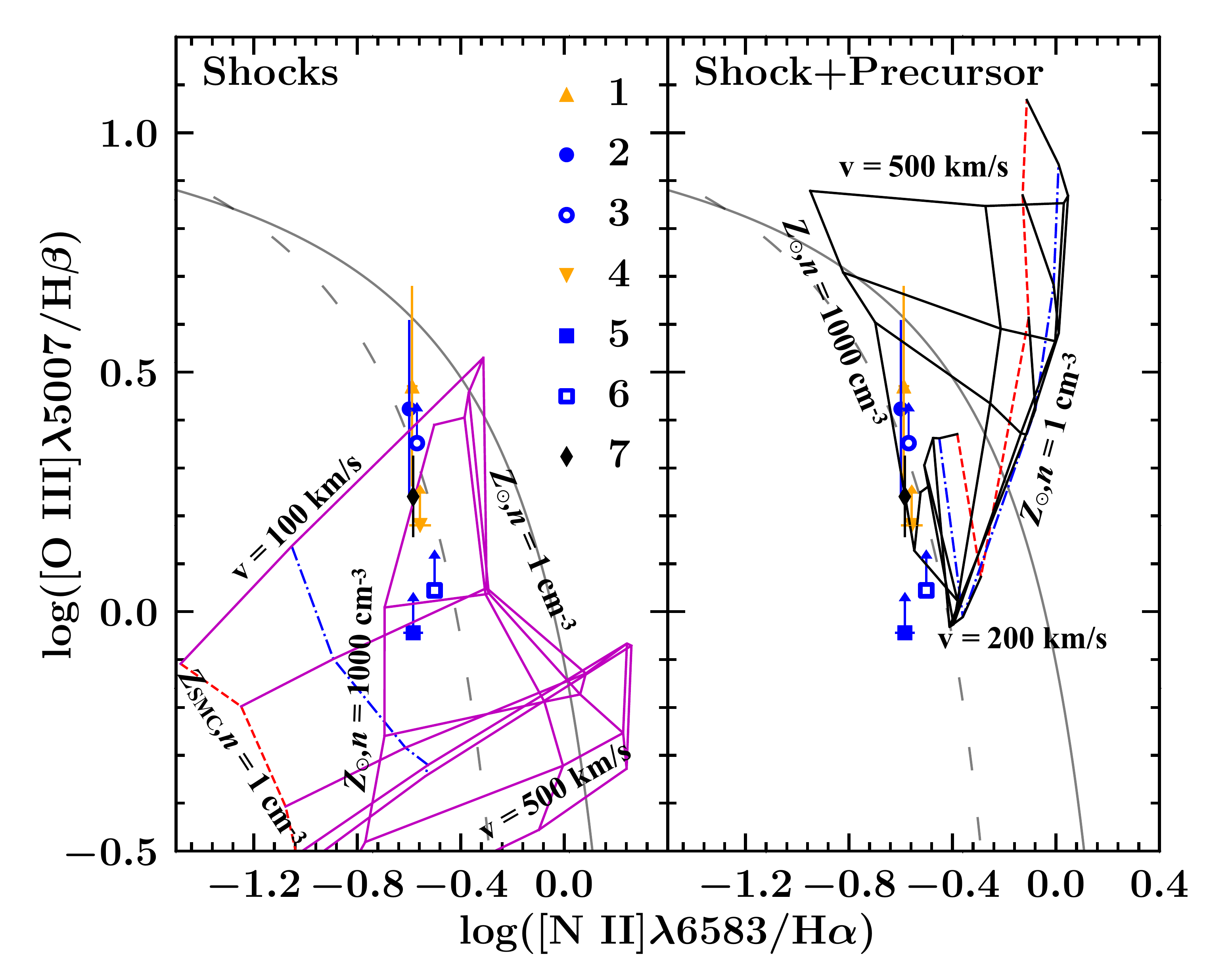}
        \includegraphics[width=0.48\textwidth]{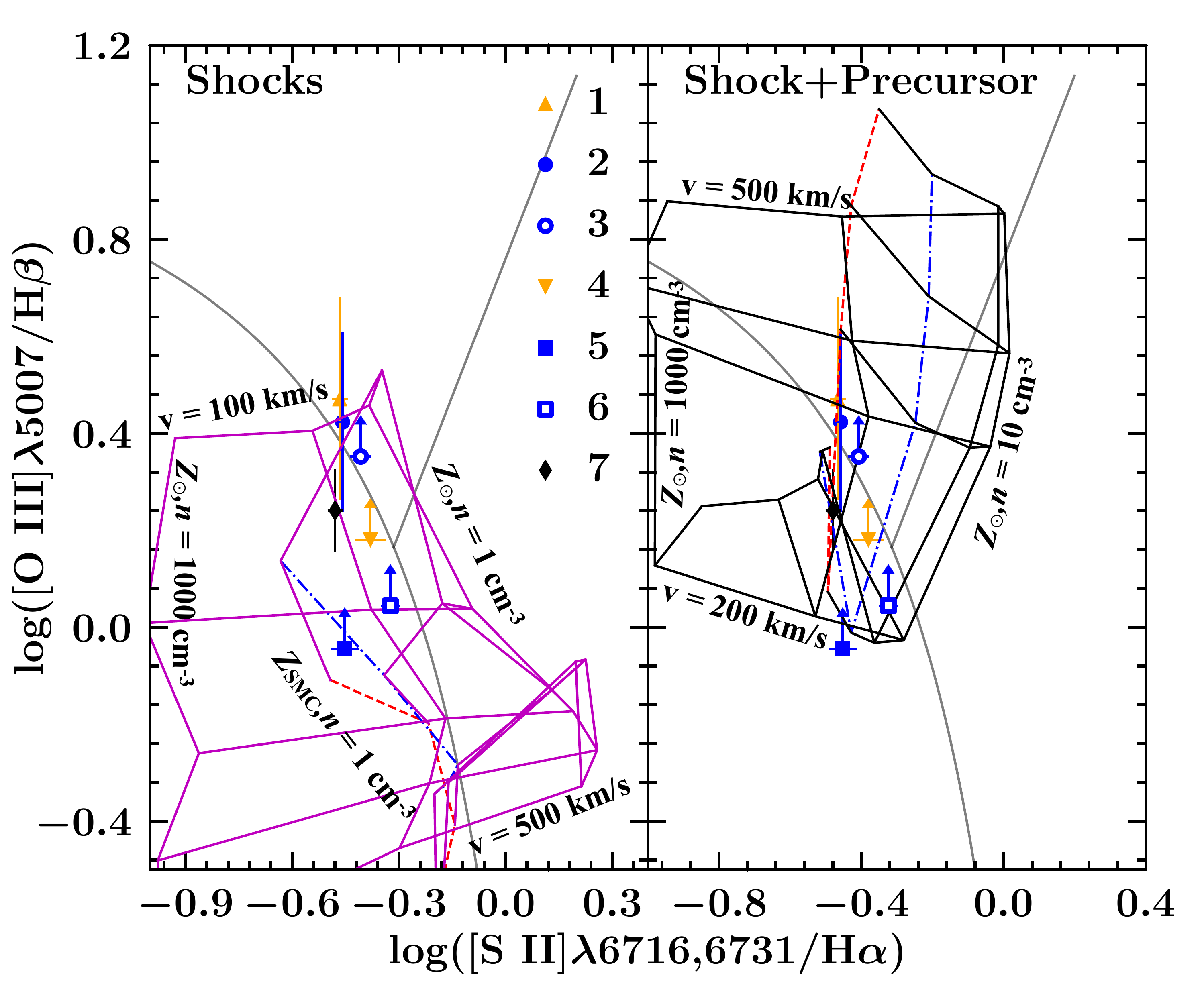}
 \includegraphics[width=0.48\textwidth]{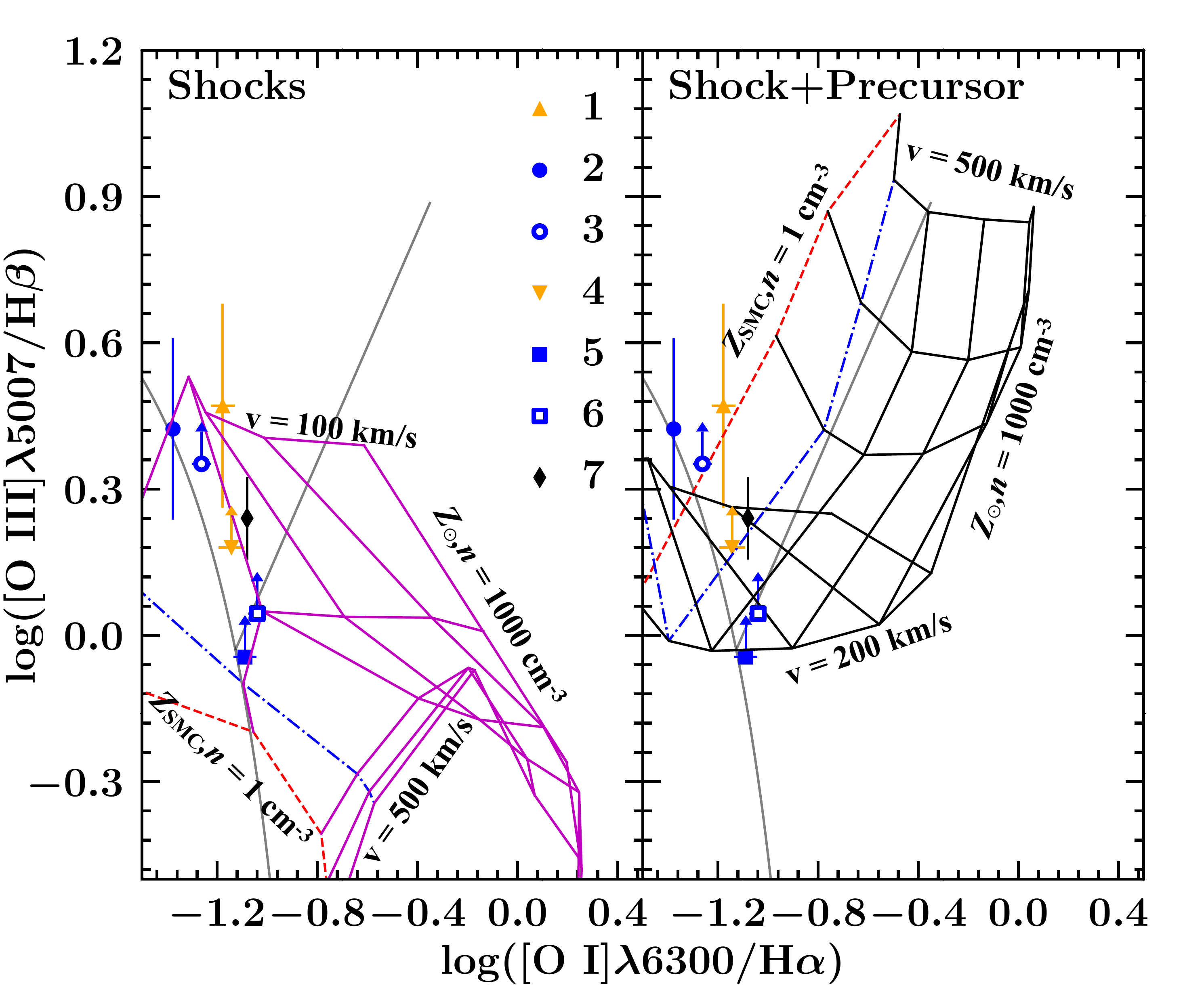}
\caption{Comparison of the emission line measurements from method 2 to the \cite{allen2008} shock and shock+precursor models. For all diagrams, the simple shock models are shown in magenta and the shock+precursor models are shown in black. The red dashed lines for both the shock and shock+precursor models represent the models assuming the SMC metallicity, while the blue dot-dashed line assumes the LMC metallicity. Both the SMC and LMC models assume a density of $n=1$~cm$^{-3}$. The edges of the grids are labeled by the value that defines the line (velocity or density). We note that the velocity always increases downwards in the shock models and upwards in the shock+precursor models. We also show standard hot star photoionization diagnostics \citep[i.e.,][]{kewley2006,Kauffmann2003} in gray. The measurements presented are from method 2, and the symbol representing each aperture is indicated in the legend in each subplot. We find the simple shocks model does a better job at describing the regions outside the radio source (all but aperture 7), and the shocks+precursor models best describe the emission coincident with the radio source (aperture 7). A detailed description of the results is given in Section~\ref{sssec:shockmod}. {\it Top Left: }The [\ion{N}{2}]]/H$\alpha$ diagram. We expect this diagram to have the most inconsistency between the data and models due to the strong dependence on metallicity. However, we find that the shocks and shocks+precursor models still do an adequate job of describing the data. {\it Top Right: }The [\ion{S}{2}]/H$\alpha$ diagram. Both the shocks and shock+precursor models describe the data well. {\it Bottom: }The [\ion{O}{1}]/H$\alpha$ diagram. The simple shocks models describe the data in aperture 2 better than the shocks+precursor models. The remaining apertures outside the radio source (aperture 1, 3, 5 and 6) can be explained by either the shocks or shocks+precursor models. While, the shocks+precursor and shocks models can both describe the emission coincident with the radio source (aperture 7), the presence of the broad H$\alpha$ and [\ion{Fe}{10}] line strongly indicates the influence of photoionization. We therefore conclude that the shocks+precursor model best describes the emission coincident with the radio source.}
    \label{fig:shock_mods}
\end{figure*}

We over-plot our data on the shock and shock+precursor models for all three BPT diagrams in Figure~\ref{fig:shock_mods}. We note that given the inconsistencies in density and metallicity between the data and the models, the [\ion{N}{2}]/H$\alpha$ diagram will likely have more systematic errors than the other two diagrams.

We only show the apertures from method 2 in these plots as method 1 is focused on integrated light. We qualitatively discuss each aperture in detail below, and give more quantitative, overall conclusions in Section~\ref{ssec:gen_con}. 

For reference, all of the apertures in method 2 are spatially resolved. Aperture 7 is the 1\arcsec\ aperture around the radio source, while apertures 1 and 4 are the two non-nuclear regions that follow the semi-major axis of the enhanced [\ion{O}{1}] emission. The remaining apertures (2, 3, 5 and 6) are non-nuclear apertures not associated with the enhanced [\ion{O}{1}] emission.

\begin{itemize}
    \item[] {\it Aperture 1: }This emission is consistent with both shocks and shocks+precursor models in all three diagrams. We therefore do not discriminate between these mechanisms, but note that shocks are contributing to the observed emission.
    \item[]{\it Aperture 2: }This emission is technically consistent with both the shocks and shocks+precursor models in all three diagrams. However, the shocks model appears to better describe the data in the [\ion{O}{1}]/H$\alpha$ diagram. In order to be explained by shocks+precursor, the gas would need to have a density less than 1~cm$^{-3}$ even with a sub-solar metallicity. This is inconsistent with our density measurement of $n_e = 500-1000$~cm$^{-3}$. We therefore conclude that the gas in aperture 2 is likely dominated by shocked emission.
    \item[]{\it Aperture 3: }We cannot discriminate between the shock and shock+precursor mechanisms given the lower limit on the [\ion{O}{3}]/H$\beta$ ratio. We do note that shocks are contributing to the observed emission.
    \item[]{\it Aperture 4: }Similar to aperture 1, we cannot discriminate between the shock and shock+precursor mechanisms but note that shocks are contributing to the observed emission.
    \item[]{\it Aperture 5: }Similar to aperture 3,  we cannot discriminate between the shock and shock+precursor mechanisms given the lower limit on the [\ion{O}{3}]/H$\beta$ ratio. We do note that shocks are contributing to the observed emission.
    \item[]{\it Aperture 6: }Similar to aperture 5, we cannot discriminate between the shock and shock+precursor mechanisms given the lower limit on the [\ion{O}{3}]/H$\beta$ ratio. We do note that shocks are contributing to the observed emission.
    \item[]{\it Aperture 7: }This aperture is centered on the radio source, and is consistent with the shocks and shock+precursor models in all three diagrams. However given the detected AGN coronal line [\ion{Fe}{10}]$\lambda$~6374, as well as the clearly broad H$\alpha$ component (${\rm FWHM}\approx900~$km~s$^{-1}$, not included in the analysis here), we know that photoionization is likely contributing to the observed emission. We therefore conclude that the mechanism that best describes the gas is a shocks+precursor model. 
\end{itemize}

\subsection{General Conclusions from Model Comparisons}\label{ssec:gen_con}
We find overall that photoionization from hot stars do not adequately describe the data. After including careful constraints on the density, metallicities, and velocities of the shock and shock+precursor models, we find that the apertures associated with the radio source and at the two edges of the enhanced [\ion{O}{1}] feature are well-described by shock+precursor models, while the other apertures are well-described by the simple shocks model. The predicted shock velocities from the model are $v\sim100-300$~km~s$^{-1}$, which is consistent measured FWHM values for the narrow-line emission. While we cannot cleanly define the predicted metallicity and density measurements expected from the models, the majority of the data lie are consistent with $n_e=10-1000$~cm$^3$ with $Z=Z_\odot$. Therefore, it is possible for the data to be consistent with $Z<Z_\odot$ and $100<n_e<1000$~cm$^3$, which agrees with the metallicity and density estimates from the nebular emission. This is especially true in the [\ion{O}{3}]/H$\beta$ vs.~[\ion{S}{2}]/H$\alpha$ diagram, which is sensitive to shock excitation \citep{Rich2010,Molina2018}. We therefore conclude that the data are consistent with the shock models. 

\section{The BH and its Environment}
\label{sec:discussion}
We conclude that an active BH is present in J1220+3020, due to the strong radio detection from \cite{reines2020}, the broad H$\alpha$ emission, enhanced [\ion{O}{1}] emission and the detection of the AGN coronal line [\ion{Fe}{10}]$\lambda$6374. In order to calculate the BH mass, we assume that all of the broad H$\alpha$ emission originates from the broad line region of the BH. Under that assumption, we use equation~5 from \cite{reines2013} and find $M_{\rm BH}=10^{4.9}M_\odot$. The predicted $M_{\rm BH}$ for J1220+3020 given its stellar mass ($\log[M_*/M_\odot]=9.4$) is $M_{\rm BH}=10^{5.5}$--$10^{6}M_\odot$ \citep{reinesvolonteri2015}. The median $M_{BH}$ in the dwarf galaxies studied in \citet{reines2013} was $M_{BH}\approx2\times10^5M_{\odot}$. Therefore our BH mass measurement is consistent with other BHs in galaxies of a similar mass range.

\subsection{Contribution of Shocks}
Despite the clear indicators described in Section~\ref{ssec:gen_con}, the traditional BPT diagnostics did not identify the object as ``AGN-like'' in all three diagrams for SDSS data and the [\ion{N}{2}]/H$\alpha$ and [\ion{S}{2}]/H$\alpha$ diagrams for GMOS. The difference between these two data sets can in part be explained by the poorer spectral resolution and larger observing region of the SDSS data. The larger SDSS aperture lets in more of the host galaxy light which will likely contaminate the AGN signal \citep{Moran2002}. Additionally, poorer spectral resolution blurs line profiles, and could potentially hide the weaker broad components, including those seen in [\ion{O}{1}]$\lambda$~6300.

In order to explain the ambiguity in the GMOS data alone, we must consider the physical structure of and power sources associated with the accreting BH. When the \cite{allen2008} shock models are included, we find that a combination of collisional shock excitation and photoionization best describes the emission coincident with the radio source. This combination of power sources could be an indicator of AGN activity, as LINERs with LLAGNs have similar optical signatures \citep{Dopita1997,Sabra2003,Molina2018}. Indeed, the fact that AGNs influence their host galaxies in ways other than photoionization is well-known \citep[e.g.,][]{Cecil1995,Capetti1997,Falcke1998,Ferruit1999,Dopita2002}, but these secondary processes are often ignored due to the dominance of photoionization \citep{Laor1998}. 

Furthermore, traditional optical AGN photoionization signatures can be overwhelmed by shock excitation from jets and other outflows from the central engine in LLAGNs with significantly larger BHs, and thus higher accretion rates \citep{Molina2018}. This effect occurs at the $100$~pc scale, which is five times smaller than the $500$~pc scale studied here. Therefore, we should expect a stronger contribution from the shocked emission relative to photoionization, given both the difference in BH mass and observed spatial scale. 

Finally the velocities predicted by the \cite{allen2008} models and the narrow emission line widths agree, indicating that the observed emission is driven in part by low-velocity shocks. The \citet{allen2008} models demonstrate that low-velocity shocks can overlap with the star-forming locus in the [\ion{N}{2}]/H$\alpha$ and [\ion{S}{2}]/H$\alpha$ diagrams, which best explains the ambiguity seen in the traditional BPT diagnostics. 

We conclude that the best-fitting physical model is a LLAGN that photoionizes the nearby gas, and drives a wind or other outflow that creates shocked emission on larger spatial scales. Outside of the outflow's direct influence, turbulent motion of the gas caused by the outflow could also create shocked emission. This model explains the enhanced [\ion{O}{1}] emission in both the 2--D map and 1--D radio source spectrum, the [\ion{Fe}{10}] and broad H$\alpha$ emission and the observed narrow emission-line ratios. While stellar winds could potentially explain the observed shock-excited gas, they cannot explain the strong [\ion{Fe}{10}] and radio emission, which strongly disfavors this scenario.

\subsection{Structure of the Central Engine}
Given the strong evidence for outflowing material shock-exciting the surrounding gas, the LLAGN structure could include a RIAF \citep{Narayan1995,Blandford1999}. Unlike the standard accretion engine, the accretion rate in a RIAF is so low ($L_{\rm bol}/L_{\rm Edd}\lesssim10^{-2}$) that there exists a radius, $R_t$ where the collisional cooling timescale is equal to the accretion timescale. Outside of $R_t$, the accretion structure is a geometrically thin, optically thick disk. Meanwhile inside $R_t$, the accretion disk becomes geometrically thick and optically thin, as the ions remain at the virial temperature while the electrons are cooled by bremsstrahlung, synchrotron, and Compton up-scattering. In fact, this structure could remove some of the more traditional AGN-like narrow-line ratios \citep{Ho2008,Trump2011}. RIAFs are both theoretically predicted and observationally shown to have strong radio outflows \citep{Meier2001,Fender2000}. Additionally, increased radio strength has been observationally tied to decreased accretion strength \citep{Melendez2008,diamondstanic2009}. 

We estimate the bolometric luminosity of the AGN by first converting the broad H$\alpha$ luminosity detected in the resolved 1\arcsec spectrum of the radio source to $L(5100$\AA) using equation 2 from \cite{Greene2005}, and then applying the $L_{\rm bol}=10.3L(5100$\AA) relation from \cite{Richards2006}. The Eddington ratio for the BH in J1220+3020 is $L_{\rm bol}/L_{\rm Edd}\sim0.03$, which is consistent with a RIAF-powered engine.

We note that there are usually no broad emission lines in RIAFs due to the cooler inner disk \citep{Trump2011}, unlike J1220+3020. However, we only detect weak broad H$\alpha$ emission, with $L({\rm H}\alpha_{\rm broad})= 4.8\times10^{38}$~erg~s$^{-1}$, and there is a precedent for detected broad emission in AGNs with very low accretion rates \citep[$L_{\rm int}/L_{\rm Edd} < 10^{-3}$; ][]{Ho2009}. 

We also note that there is a $\sim0.5$~dex uncertainty in virial $M_{\rm BH}$ estimates \citep{vestergaard2006,Shen2013}, including the one quoted here. Therefore, the range of $M_{\rm BH}=10^{4.4}$--$10^{5.4}M_\odot$ results in an Eddington ratio within the range $\sim0.001$--0.1, where the lower Eddington ratio corresponds to the higher $M_{\rm BH}$ estimate. If we have over-estimated the BH mass, then the accretion rate would be too high to be explained by a RIAF engine. Therefore, without a full SED, we cannot definitively claim the central engine has a RIAF. However, the lack of AGN-like narrow emission-line ratios, the strong radio emission and the evidence for AGN outflows, i.e., [\ion{Fe}{10}] and enhanced [\ion{O}{1}], are all consistent with the RIAF model, and we therefore favor that explanation.

If there is a RIAF, the BH mass and AGN luminosity estimates could be impacted by the potential suppression of broad emission by the cooler inner disk \citep{yuan2004,Ho2008,Trump2011}, but we reiterate that our estimated BH mass is consistent with those in the same galaxy mass range.

\subsubsection{Contribution from the Radio Jet}

We use the VLA Faint Images of the Radio Sky at Twenty centimeters (FIRST) survey 1.4~GHz detection \citep{Becker1995} and equation 1 from \cite{cavagnolo2010} to calculate the jet power. Given that \cite{reines2020} found that the radio emission is dominated by the AGN candidate in this object, we assume there is little contamination by star formation. We note that this relation was calculated using AGNs in giant ellipticals with radio luminosities similar to that of J1220+2030. If we assume this relation holds, we calculate $P_{jet}\approx8.3\times10^{41}$~erg~s$^{-1}$, which is about 9\% of the Eddington luminosity assuming the BH mass given above. Thus the majority of AGN power is likely driven by emission from the radio jet, not accretion.

As discussed in \cite{cavagnolo2010}, the $P_{jet}$ presented here is actually a calculation for $P_{cav}$, which is the mechanical power needed to create the radio cavity. However, they note that since the mechanical $P_{cav}$ is significantly larger than the synchrotron power of the jet, they assume $P_{jet}=P_{cav}$. Finally, this measurement does not include the energy channeled into shocks, which can greatly exceed $P_{cav}$ \citep{Nulsen2005}.

In addition to the low Eddington ratio, the RIAF model is associated with very strong radio emission and evidence for outflows or other turbulent motion in the gas creating shocked emission. We see evidence for very strong radio emission via the larger energy contribution from the radio jet than accretion. While we do not observe large radio outflows, this black hole has a $M_{\rm BH}=10^{4.9}M_{\odot}$, which would likely not be able to produce large radio lobes. However, there is evidence for turbulent motion and outflows via the [\ion{Fe}{10}] emission. While the coronal [\ion{Fe}{10}] line is typically thought to be powered by AGN photoionization \citep[e.g.,][]{Nussbaumer1970,Korista1989,Oliva1994,Pier1995}, it can also be produced by shock-excited gas from out-flowing winds caused by radio jets \citep{wilson1999}. If we assume the latter scenario, we can calculate the expected velocity of the jet-driven shock using the [\ion{Fe}{10}]/H$\beta$ ratio. The predicted velocity is $\sim200-300$~km~s$^{-1}$, which is consistent with the measured line dispersions (${\rm FWHM}\sim100-300$~km~s$^{-1}$). Given the narrow-line evidence for shocks and the radio detection from \cite{reines2020}, we conclude the [\ion{Fe}{10}] line is likely created by shocks driven by out-flowing winds from the radio jets associated with the BH.

\subsection{Comparison to Optically-Selected BHs}
We conclude that the engine in J1220+3020 is largely consistent with the RIAF model, and that the shocked emission due to out-flowing gas or winds partially obscures the optical AGN photoionization signatures. This indicates that radio-selected AGN may probe a different population of active BHs in dwarf galaxies than standard optical diagnostics. 

Optical diagnostics are known to pick out bright, high accretion- rate BHs in dwarf galaxies where photoionization dominates the energy budget \citep{reines2016}. As the accretion rate decreases, radio emission increases and outflows and winds become more important to the overall energy budget. Therefore, radio-selected BHs in dwarf galaxies will likely have lower accretion rates than those found in optical surveys. Given that a significant fraction of supermassive active BHs in the local Universe are LLAGNs \citep{Ho2008}, radio surveys might find a significant population of BHs in local dwarf galaxies.  
\section{Summary and Conclusions}
\label{sec:summary}

In this paper we present the first optical follow-up of the radio-selected BH candidates discovered by \cite{reines2020}. We focus on J1220+3020, which is a bright nuclear BH candidate, to identify any unique features associated with radio-selected AGNs in dwarf galaxies. We use GMOS-N/IFU data to study look for optical signatures of an active BH, and study its 2--D physical environment. Our results are as follows:

\begin{itemize}
    \item There is an elongated feature of enhanced [\ion{O}{1}]$\lambda$6300 emission centered on the radio source and a broad (FWHM $\approx 700$ km s$^{-1}$) [\ion{O}{1}] component in the 1--D spectrum of the radio source. We conclude these are likely signatures of an outflow.
    
    \item In the 1--D spectrum of the radio source, we detect the AGN coronal line [\ion{Fe}{10}]$\lambda$6374 and broad H$\alpha$ (${\rm FWHM}\approx900$~km~s$^{-1}$), which both indicate the presence of an AGN. Using the FWHM and luminoisity of the broad H$\alpha$ emission, we estimate a BH mass of $M_{\rm BH}=10^{4.9}M_\odot$.
    
    \item We compare our data to both standard BPT diagnostics and the \cite{allen2008} shock and shock+precursor models. We conclude that the LLAGN photoionizes the gas in its immediate vicinity, while an outflow or winds associated with the LLAGN shock-excites gas further away and drives the [\ion{Fe}{10}]$\lambda$6374 emission.
    
    \item We favor the RIAF model as it is consistent with almost all of the observed data. Finally, we conclude that shocks associated with outflows or winds significantly contribute to the AGN energy budget.
    
    \item Sensitive, high resolution radio surveys of dwarf galaxies can probe a population of BHs with lower accretion rates than optical surveys.
\end{itemize}

While we do not find any 2--D velocity structure that would indicate an an outflow, the enhanced 2--D [\ion{O}{1}] emission and broadened [\ion{O}{1}] emission associated with that feature, the strong radio and [\ion{Fe}{10}] detection, and the agreement of shock models with the [\ion{Fe}{10}]/H$\beta$ ratio and the measured forbidden emission-line widths all strongly support this scenario. Finally, we note the AGN in J1220+3020 would not be identified through the [\ion{N}{2}]/H$\alpha$ diagram alone, nor relying solely on standard BPT diagnostics. Therefore by limiting our work to that diagram and set of models, we are likely missing a large portion of the AGN population in dwarf galaxies. In order to address this problem, future work should include a the use of shock models, and if possible [\ion{O}{1}], when exploring LLAGNs in dwarf galaxies.  

\acknowledgements
We thank the anonymous referee for insightful comments that helped us improve this paper. AER acknowledges support for this work provided by Montana State University and NASA through EPSCoR grant number 80NSSC20M0231.
This work is based on observations obtained through program GN-2020A-Q-215 at the international Gemini Observatory, a program of NSF’s NOIRLab, acquired through the Gemini Observatory Archive at NSF’s NOIRLab and processed using the Gemini IRAF package, which is managed by the Association of Universities for Research in Astronomy (AURA) under a cooperative agreement with the National Science Foundation on behalf of the Gemini Observatory partnership: the National Science Foundation (United States), National Research Council (Canada), Agencia Nacional de Investigaci\'{o}n y Desarrollo (Chile), Ministerio de Ciencia, Tecnolog\'{i}a e Innovaci\'{o}n (Argentina), Minist\'{e}rio da Ci\^{e}ncia, Tecnologia, Inova\c{c}\~{o}es e Comunica\c{c}\~{o}es (Brazil), and Korea Astronomy and Space Science Institute (Republic of Korea). This work was enabled by observations made from the Gemini North telescope, located within the Maunakea Science Reserve and adjacent to the summit of Maunakea. We are grateful for the privilege of observing the Universe from a place that is unique in both its astronomical quality and its cultural significance.

Funding for the Sloan Digital Sky Survey IV has been provided by the Alfred P. Sloan Foundation, the U.S. Department of Energy Office of Science, and the Participating Institutions. SDSS-IV acknowledges support and resources from the Center for High Performance Computing  at the University of Utah. The SDSS website is www.sdss.org.

SDSS-IV is managed by the Astrophysical Research Consortium for the Participating Institutions of the SDSS Collaboration including the Brazilian Participation Group, the Carnegie Institution for Science, Carnegie Mellon University, Center for Astrophysics | Harvard \& Smithsonian, the Chilean Participation Group, the French Participation Group, Instituto de Astrof\'isica de Canarias, The Johns Hopkins University, Kavli Institute for the Physics and Mathematics of the Universe (IPMU) / University of Tokyo, the Korean Participation Group, Lawrence Berkeley National Laboratory, Leibniz Institut f\"ur Astrophysik Potsdam (AIP),  Max-Planck-Institut f\"ur Astronomie (MPIA Heidelberg), Max-Planck-Institut f\"ur Astrophysik (MPA Garching), Max-Planck-Institut f\"ur Extraterrestrische Physik (MPE), National Astronomical Observatories of China, New Mexico State University, New York University, University of Notre Dame, Observat\'ario Nacional / MCTI, The Ohio State University, Pennsylvania State University, Shanghai Astronomical Observatory, United Kingdom Participation Group, Universidad Nacional Aut\'onoma de M\'exico, University of Arizona, University of Colorado Boulder, University of Oxford, University of Portsmouth, University of Utah, University of Virginia, University of Washington, University of Wisconsin, Vanderbilt University, and Yale University.

 This research made use of Astropy,\footnote{http://www.astropy.org} a community-developed core Python package for Astronomy \citep{astropy2013, astropy2018}. 

\software{APLpy \citep{aplpy},
Astropy \citep{astropy2013,astropy2018},
Gemini IRAF \citep{gmosiraf},
IFSRED \citep{gmosifu},
IRAF \citep{iraf},
L.A.-cosmic \citep{vanDokkum2001},
Matplotlib \citep{matplotlib},
PyRAF \citep{pyraf},
pyspeckit \citep{pyspeckit}}

\appendix

\section{Emission-Line Fits for the 1\arcsec\ Radio Source Spectrum}\label{app:apb_info}
\begin{figure}[t]
    \centering
    \includegraphics[width=0.46\textwidth]{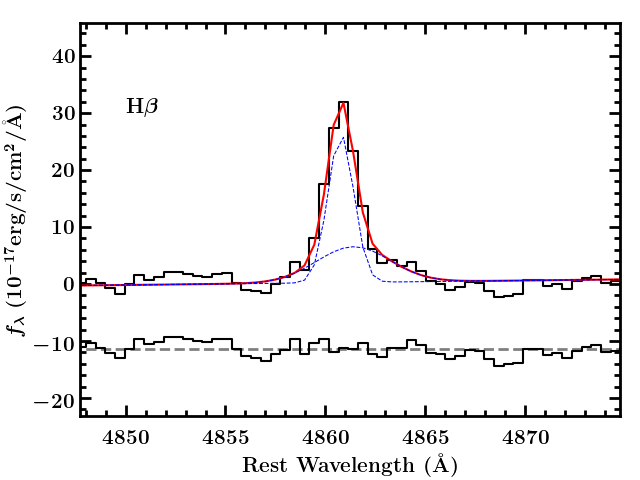}
    \includegraphics[width=0.46\textwidth]{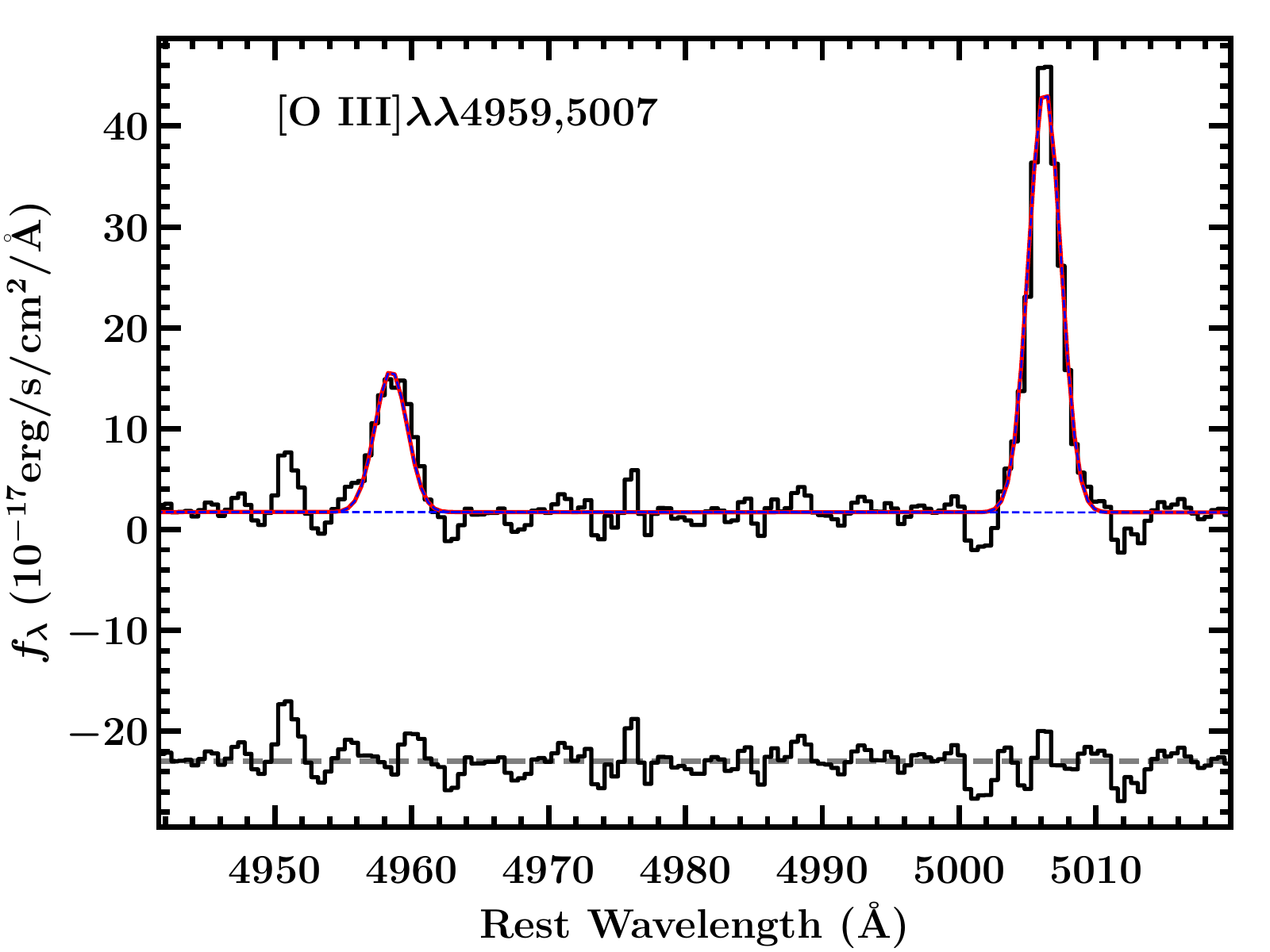}
    \includegraphics[width=0.48\textwidth]{o1_broad_v2.pdf}
    \includegraphics[width=0.46\textwidth]{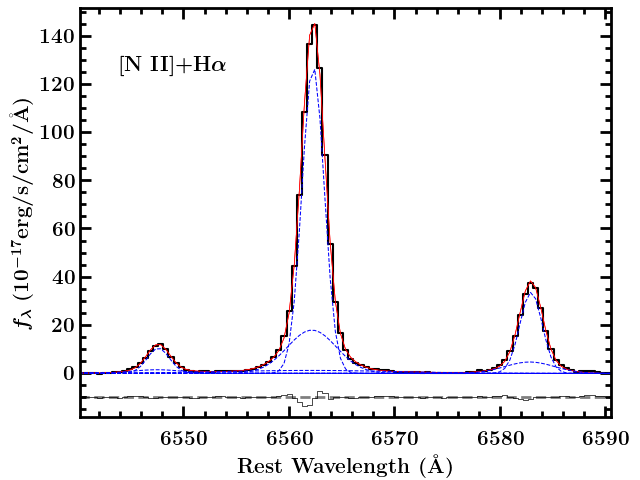}
     \includegraphics[width=0.46\textwidth]{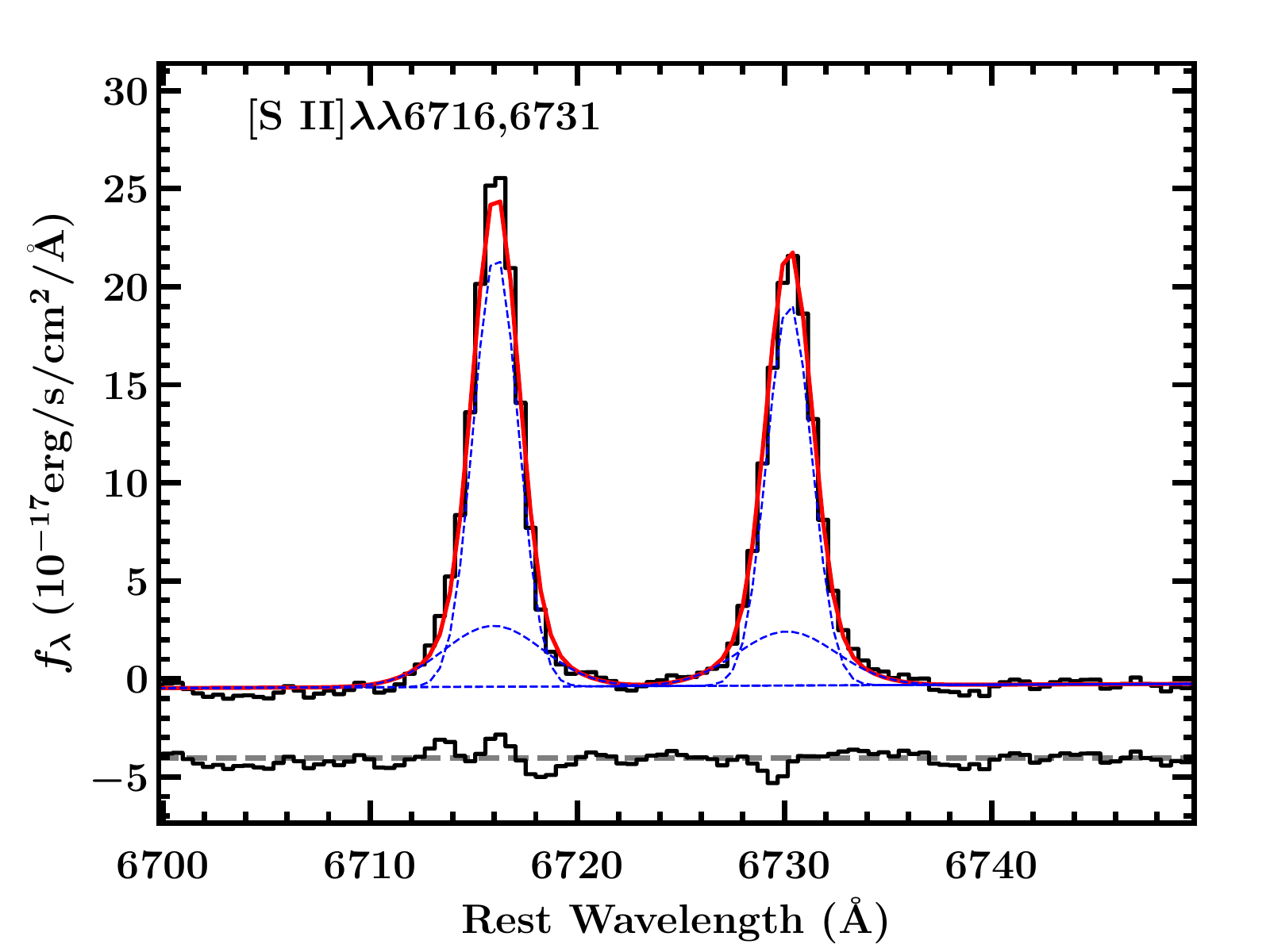}
\caption{The fits for the emission lines in the 1\arcsec\ spectrum of the radio source. In each plot, the fit is shown in red, the data are shown in black and the individual components are shown in blue. The emission lines shown are indicated in each subplot. The broad component of the [\ion{O}{1}] component is shown in orange. We do not do the same for the broad component of H$\alpha$, as it is not clearly visible in this image. We show a zoomed in plot of the H$\alpha$ broad feature in Figure~\ref{fig:habroad}.}
    \label{fig:rs_fits}
\end{figure}

In this appendix we show the fits to the strong emission lines in the 1\arcsec\ spectrum of the radio source in Figure~\ref{fig:rs_fits}. We specifically focus on the emission lines used in our analysis.

\clearpage

\bibliographystyle{apj}
\bibliography{gemini.bib}

\clearpage

\movetabledown=2in
\begin{rotatetable}
\begin{deluxetable*}{ccccccccccccc}
  \tablecaption{Narrow Emission Line Fluxes for J1220+3020\label{table:fluxes}}
\tabletypesize{\footnotesize}
\setlength{\tabcolsep}{4pt}
\renewcommand{\arraystretch}{1.}
\tablewidth{0pt}
\tablehead{
\colhead{} & {} & \multicolumn{11}{c}{Emission Line Flux\tablenotemark{a}}\vspace{-4mm}\\
\colhead{} & {} & \multicolumn{11}{c}{\hrulefill}\\
\colhead{Method} &{Number} & {H$\beta$~$\lambda4861$} & {[\ion{O}{3}]$\lambda$4959} & {[\ion{O}{3}]$\lambda$5007} & {[\ion{O}{1}]$\lambda$6300} & {[\ion{O}{1}]$\lambda$6363} & {[\ion{Fe}{10}]$\lambda$6364} & {[\ion{N}{2}]$\lambda$6548} & {H$\alpha$~$\lambda$5007} & {[\ion{N}{2}]$\lambda$6583} & {[\ion{S}{2}]$\lambda$6716} & {[\ion{S}{2}]$\lambda$6731}}
\startdata
{1} & {A\tablenotemark{b}} & {$24\pm3$} & {$14\pm1$} & {$43\pm2$} & {$14\pm1$} & {$4.7\pm0.3$} & {$1.1\pm$0.4} & {$13\pm1$} & {$148\pm5$} & {$39\pm2$} & {$25\pm4$} & {$23\pm4$} \\
{1} & {B} & {$70\pm10$} & {$41\pm1$} & {$124\pm4$} & {$37\pm2$} & {$12\pm1$} & {$3.0\pm0.7$} & {$38.7\pm0.1$} & {$445\pm1$} & {$116.1\pm0.4$} & {$78\pm5$} & {$69\pm5$}\\
{1} & {C} & {$110\pm10$} & {$70\pm2$} & {$211\pm5$} & {$71\pm3$} & {$23\pm1$} & {...} & {$74\pm0.3$} & {$808\pm2$} & {$222\pm1$} & {$160\pm10$} & {$134\pm9$}\\
{2} & {1} & {$6\pm3$} & {$6\pm1$} & {$19\pm2$} & {$5.5\pm0.6$} & {$1.8\pm0.2$} & {...} & {$7.2\pm0.1$} & {$83\pm4$} & {$21.6\pm0.3$} & {$14.8\pm0.7$} & {$13.7\pm0.5$} \\
{2} & {2} & {$10\pm4$} & {$9\pm1$} & {$27\pm2$} & {$4.7\pm0.3$} & {$1.6\pm0.1$} & {...} & {$9.4\pm0.1$} & {$113\pm5$} & {$28.3\pm0.4$} & {$20.9\pm0.7$} & {$18.3\pm0.5$} \\
{2} & {3} & {$<10$} & {$9\pm1$} & {$27\pm2$} & {$5.2\pm0.3$} & {$1.7\pm0.1$} & {...} & {$8.6\pm0.1$} & {$95\pm6$} & {$25\pm0.4$} & {$21.0\pm0.5$} & {$16.4\pm0.4$} \\
{2} & {4} & {$<10$} & {$5\pm1$} & {$15\pm2$} & {3.4$\pm0.2$} & {$1.1\pm0.1$} & {...} & {$4.3\pm.1$} & {$46\pm4$} & {$12.9\pm0.3$} & {$10.7\pm0.4$} & {$8.7\pm0.3$} \\
{2} & {5} & {$<10$} & {$2.7\pm0.3$} & {$8\pm1$} & {$3.0\pm0.2$} & {$1.0\pm0.1$} & {...} & {$3.2\pm0.1$} & {$37\pm3$} & {$9.7\pm0.2$} & {$6.8\pm0.3$} & {$6.4\pm0.3$}\\
{2} & {6} & {$<10$} & {$5\pm1$} & {$16\pm2$} & {$4\pm0.3$} & {$1\pm0.1$} & {...} & {$4.6\pm0.1$} & {$44\pm2$} & {$13.9\pm0.3$} & {$11.3\pm0.6$} & {$9.6\pm0.4$} \\
{2} & {7\tablenotemark{c}} & {$70\pm10$} & {$41\pm1$} & {$124\pm4$} & {$37\pm2$} & {$12\pm1$} & {$3.0\pm0.7$} & {$38.7\pm0.1$} & {$445\pm1$} & {$116.1\pm0.4$} & {$78\pm5$} & {$69\pm5$}\\
{N/A} & {SDSS\tablenotemark{d}} & {$320\pm20$} & {$130\pm10$} & {$400\pm30$} & {$68\pm6$} & {$23\pm2$} & {...} & {$101\pm5$} & {$1150\pm50$} & {$299\pm7$} & {$250\pm10$} & {$182\pm7$}\\
\enddata
\tablenotetext{a}{Fluxes are presented in units of $10^{-17}$~erg~s$^{-1}$~cm$^{-2}$, and are not corrected for reddening.}
\tablenotetext{b}{Method 1, aperture A is not spatially resolved.}
\tablenotetext{c}{This Spectrum is the same as method 1, aperture B.}
\tablenotetext{d}{The SDSS measurements presented here were calculated by \cite{reines2020}, and use the 3\arcsec SDSS DR8 spectrum.}
\end{deluxetable*}
\end{rotatetable}

\clearpage
\label{lastpage}
\end{document}